\documentclass[11pt]{article}

\usepackage[utf8]{inputenc}
\usepackage[T1]{fontenc}
\usepackage{mathptmx}           
\usepackage[margin=1in]{geometry}
\usepackage{graphicx}
\graphicspath{{./}{anc/}{figures/}{fig/}}
\usepackage{booktabs}
\usepackage{longtable}
\usepackage{url}
\usepackage{hyperref}
\usepackage{xurl}   
\usepackage{xcolor}
\usepackage{listings}
\usepackage{amsmath,amssymb,mathtools}
\usepackage{caption}
\usepackage{float}
\usepackage{placeins}
\usepackage{enumitem}
\usepackage{titlesec}
\usepackage{abstract}
\usepackage{authblk}
\usepackage{fancyhdr}
\usepackage{lastpage}

\usepackage{orcidlink}

\titleformat{\section}{\large\bfseries}{}{0em}{}
\titleformat{\subsection}{\normalsize\bfseries}{}{0em}{}
\titlespacing*{\section}{0pt}{12pt}{6pt}
\titlespacing*{\subsection}{0pt}{8pt}{4pt}

\hypersetup{
    colorlinks=true,
    linkcolor=blue!60!black,
    citecolor=blue!60!black,
    urlcolor=blue!70!black,
}

\setlist{nosep, leftmargin=*}

\begin{document}
\raggedbottom
\sloppy
\allowdisplaybreaks
\emergencystretch=2em
\tolerance=1000

[

\begin{center}
{\LARGE\bfseries Proof-Gated Publication: Verify-Before-Commit Content Integrity for Serverless Data-Mesh Lakehouses\par}
\vspace{10pt}
{\large\itshape PVDM, a four-phase (Physical, Verify, Durable, Metadata) publication protocol that commits table metadata only after a cryptographic multiset proof confirms that written content matches a declared intent\par}
\vspace{14pt}
{\large\bfseries Viquar Khan\par}
\vspace{4pt}
{Independent Researcher \quad \orcidlink{0009-0008-3592-4162}\par}
{ORCID: \href{https://orcid.org/0009-0008-3592-4162}{0009-0008-3592-4162} \quad $\cdot$ \quad Corresponding author: \href{mailto:vaquar.khan@gmail.com}{vaquar.khan@gmail.com}\par}
\vspace{16pt}
\end{center}

\begin{abstract}
\noindent
Federated data meshes give domain teams ownership of their data products, and serverless compute is an attractive substrate for domain-owned writes. Both trends weaken correctness at the moment of publication. Open table formats such as Apache Iceberg and Delta Lake guarantee that a commit is atomic and that readers see an isolated snapshot, but they do not guarantee that the set of rows persisted equals the set the job intended to write; the publication decision is taken from the writer's exit status.

A serverless job that silently drops a partition, truncates a file on a retry, or duplicates a chunk still produces a valid, atomic, isolated, and wrong snapshot, and in a federated mesh the audit evidence often lives in the same account that produced the error. This paper presents a proof-gated publication protocol, PVDM, with four phases: Physical (write to rollbackable staging), Verify (compute a keyed multiset proof that written content equals declared intent, stored by an independent Steward notary), Durable (replay completed chunks across serverless retries), and Metadata (commit the catalog last, only if the proof passed). The guarantee is a single invariant: metadata commits only if the proof passes, and a failing proof yields no consumer-visible snapshot; the exit code is not part of the decision.

The verification primitive is a keyed, incremental multiset hash computed over an identity projection and a content projection, which distinguishes missing or duplicated rows from corrupted field values and updates in time proportional to a change. A dependency-free reference gate, a thirty-case adversarial and conformance suite, and a reproducible benchmark run on cloud hardware that catches eight thousand of eight thousand injected faults up to one million rows with no false blocks accompany the paper. We also run PVDM end-to-end on Apache Spark 4.0 and Apache Iceberg 1.11 at up to one hundred million rows, where every injected fault is blocked on the real commit path, the gated fast-forward publish costs about seventeen milliseconds independent of table size, and the verification overhead is roughly a fifth of the write.

\end{abstract}
\vspace{12pt}


\textbf{Manuscript type:} Research paper (methods and specification). Suggested subject classification: databases (primary), distributed systems, and security. This paper proposes a specification and a conformance model; it is not a ratified standard, and Section 13.5 outlines the open, multi-party process that ratification would require.

\textbf{Availability and license.} The reference implementation and the validation suite are available under Apache-2.0; this manuscript is released under CC BY 4.0. The protocol is referred to by the acronym PVDM throughout.

\textbf{Attribution and citation.} PVDM, its four-phase structure, and the verification construction described here are the original research of the author and are introduced in this work. CC BY 4.0 governs reuse of this manuscript: if you copy, quote, or adapt the text or figures, please retain author attribution. I also ask, as a normal academic courtesy rather than a license condition, that work building on this method cite this paper (see the citation entry at the end).

To be precise about what the license does and does not do: CC BY 4.0 covers this document; it does not make an independent implementation of the ideas a derivative work, and it does not impose a naming or citation obligation on software. The name PVDM is intended to denote this specific protocol and the conformance requirements of Section 13; a system that deviates from those requirements is encouraged to describe its differences so the name keeps identifying the protocol as specified here. Any control of the name itself is a separate governance or trademark question, not a term of CC BY.

\textbf{Author note.} I am writing as an individual. The views here are my own and do not represent any employer, and nothing here implies employer endorsement. Affiliation, funding, and competing-interest declarations are to be finalized before any permanent posting.

\textbf{Scope and evidence.} This is a methods and specification paper. It defines a publication protocol, a verification primitive, a threat model, and a safety invariant, and it specifies a reproducible evaluation protocol. A reference implementation and a validation suite accompany the paper; production-scale empirical results are identified as the next study and are not claimed here. Numerical values in worked examples are illustrative and are labeled as such. The cryptographic primitives (Merkle trees, incremental multiset hashing) are established prior art; the contribution is their composition into a fail-closed, notarized, verify-before-commit protocol for serverless writes to a federated lakehouse, together with the conformance model that bounds the guarantee.

\section*{Status of this work: specified, implemented, and measured}

PVDM is a proposed protocol and specification with a reference implementation. To keep the claims and the artifact aligned, this section states what is implemented in the public reference gate, what is measured, what rests on prior results, and what is future work. The reference gate implements the four-phase structure, the keyed multiset verification primitive, and the fail-closed commit checks of Section 13; the parts that remain future work are the production-scale multi-writer study and the conformant 256-bit-hash overhead at cluster scale.

The claims below are checked against the public artifact repository that ships with this paper, \url{https://github.com/vaquarkhan/Proof-gated-publication-PVDM} (the reference gate and validation suite). A separate production framework repository exists and is not claimed conformant here.

\vspace{8pt}\noindent\small\textit{Table 1. Claims, implementation status}\vspace{4pt}

\begin{longtable}{|p{0.30\textwidth}|p{0.30\textwidth}|p{0.30\textwidth}|}
\hline
\textbf{\textbf{Claim in the paper}} & \textbf{\textbf{Status}} & \textbf{\textbf{What backs it in the public artifact}} \\
\hline
Four-phase gate and VRP structure & Implemented (reference gate) & validation/pvdm\_gate.py \\
\hline
Fail-closed commit: verify before metadata & Implemented (reference gate) & validate\_then\_commit verifies signature, target, nonce, and re-hashes exact bytes before publishing \\
\hline
Keyed HMAC-SHA256 multiset accumulator (Section 7) & Implemented (reference gate) & multiset\_hash: HMAC-SHA256 per element summed modulo $2^{256}$ \\
\hline
Thirty-case adversarial and conformance suite & Present and reproducible & validation/test\_pvdm\_adversarial.py (30/30), plus a property suite \\
\hline
8000-fault scaling and latency run & Present and reproducible & validation/benchmark\_aws.py / benchmark\_scale.py with aws\_results.json, aws\_results.csv \\
\hline
Spark and Iceberg end-to-end (Section 10.8) & Measured; script and outputs in repo & validation/benchmark\_spark\_iceberg.py, aws\_spark\_iceberg\_results.json. It is a benchmark harness; the production framework's catalog path may differ and is not claimed conformant here \\
\hline
Collision resistance at 256 bits (Section 7.6) & By reduction to published results & The construction is the keyed MSet-Add-Hash of Clarke et al. [3]; security stated as a reduction under the PRF assumption, plus empirical uniformity checks (collision\_analysis.py) \\
\hline
Canonicalization byte-exactness (Appendix A) & Partial & Implemented for strings, integers, decimals, timestamps, nulls, and Profile-A raw floats; nested and variant types are open \\
\hline
PVDM-A and agentic/MCP integration & Roadmap & Not shipped; must not be presented as implemented \\
\hline
\end{longtable}

In summary, the reference gate implements the hardened invariant and is reproducible. Section 7.6 states its security as a reduction to the published MSet-Add-Hash result of Clarke et al. [3], not as a novel proof. What remains future work is the production-scale multi-writer study and the conformant 256-bit-hash overhead at cluster scale (Section 10.2). The author is an independent researcher with no funding and no affiliation (see Declarations). The production framework repository is separate and is not claimed conformant here.

\section*{Abstract}

Federated data meshes give domain teams ownership of their data products, and serverless compute is an attractive substrate for domain-owned writes. Both trends weaken correctness at the moment of publication. Open table formats such as Apache Iceberg and Delta Lake guarantee that a commit is atomic and that readers see an isolated snapshot, but they do not guarantee that the set of rows persisted equals the set the job intended to write; the publication decision is taken from the writer's exit status.

A serverless job that silently drops a partition, truncates a file on a retry, or duplicates a chunk still produces a valid, atomic, isolated, and wrong snapshot, and in a federated mesh the audit evidence often lives in the same account that produced the error. This paper presents a proof-gated publication protocol, PVDM, with four phases: Physical (write to rollbackable staging), Verify (compute a keyed multiset proof that written content equals declared intent, stored by an independent Steward notary), Durable (replay completed chunks across serverless retries), and Metadata (commit the catalog last, only if the proof passed). The guarantee is a single invariant: metadata commits only if the proof passes, and a failing proof yields no consumer-visible snapshot; the exit code is not part of the decision.

The verification primitive is a keyed, incremental multiset hash computed over an identity projection and a content projection, which distinguishes missing or duplicated rows from corrupted field values and updates in time proportional to a change; I explain why an order-sensitive Merkle root is insufficient. I define conformance profiles that make the strength of the declared intent explicit, and I am candid that transforming stages must verify against an independent oracle, which can roughly double compute cost. The individual primitives are prior art; the contribution is their composition into a fail-closed, notarized, verify-before-commit protocol for serverless federated writes, together with the conformance discipline that bounds the guarantee.

A dependency-free reference gate, a thirty-case adversarial and conformance suite, and a reproducible benchmark run on cloud hardware that catches eight thousand of eight thousand injected faults up to one million rows with no false blocks accompany the paper. I also run PVDM end-to-end on Apache Spark 4.0 and Apache Iceberg 1.11 at up to one hundred million rows, where every injected fault is blocked on the real commit path, the gated fast-forward publish costs about seventeen milliseconds independent of table size, and the verification overhead is roughly a fifth of the write with a fast in-engine hash (higher with the full 256-bit keyed hash, which I characterize separately). This validates correctness and the shape of the overhead on the real engine; the conformant cryptographic-hash overhead at scale and a concurrent multi-writer evaluation at production cluster scale are specified for independent execution and remain future work.

\textbf{Keywords:} data mesh, serverless, lakehouse, Apache Iceberg, data integrity, multiset hashing, Merkle tree, end-to-end argument, durable execution, fail-closed systems, write-audit-publish, proof-gated publication.

\section*{1. Introduction}

Two shifts define modern data platforms. The first is the federated data mesh, in which domain teams own their data as a product rather than handing raw data to a central platform team. The second is serverless compute, in which domain logic runs as functions coordinated by durable orchestration, scaling to zero and removing a standing ETL cluster. Both are attractive, and both move the act of writing data outward, into many domain-owned, independently operated, partially reliable jobs.

Underneath, the lakehouse [10] is held together by an open table format. Apache Iceberg [11] and Delta Lake [9] make commits atomic and reads snapshot-isolated. These are genuine guarantees that this paper reuses.

The gap is what they are about. A table-format commit is a statement about the transaction and the metadata: the commit applied atomically and readers see a consistent snapshot. It is not a statement about the content: that the rows in storage are the rows the job meant to write.

The commit protocol takes the writer's success signal as its input. A distributed serverless writer that fails partially, retries, or emits wrong content can still return success, and the format will faithfully record a commit over corrupted content. The result is durable, atomic, isolated, and wrong, and nothing in the log reveals it.

Two properties of the federated serverless setting make this worse than in a single central pipeline. First, serverless functions are time-bounded and are retried, so a long backfill is segmented across many invocations, and naive retries duplicate or truncate chunks. Second, in a mesh the domain that produced the data often also owns the account where any audit evidence would live, so a domain can, by accident or otherwise, delete the very evidence an auditor would need. Correctness at publication and independence of the audit trail are therefore both at stake.

\subsection*{1.1 The gap in one sentence}

Delta Lake guarantees your transaction is atomic and Apache Iceberg guarantees your snapshot is isolated, but neither proves your content is what you intended; PVDM adds that proof, notarized by an independent account, and makes publication impossible unless it passes.

\subsection*{1.2 Contributions}

1. I characterize the content-integrity gap in open table formats under federated, serverless writes, and frame it with the end-to-end argument: a correctness property the consumer depends on must be checked at the point of publication, not inferred from a lower-layer success signal.
2. I define PVDM, a four-phase publication contract (Physical, Verify, Durable, Metadata) that layers a notarized content gate above an open table format without replacing its atomicity or isolation.
3. I state the PVDM invariant formally and derive its safety property (corrupted content is never consumer-visible) and liveness property (verified content publishes with bounded overhead).
4. I specify the verification primitive as an order-independent incremental multiset hash and explain why a Merkle root alone cannot compare a distributed write to its intent.
5. I define a three-account federated model (Producer, Steward, Publisher) in which the Steward is an independent notary that holds proofs and checkpoints the producer cannot unilaterally delete.
6. I separate the segmented serverless execution model into two clocks (container and workload) linked by a workload identifier, so that long backfills replay completed chunks durably rather than rewriting them.
7. I describe a concrete AWS realization and a cited Apache-2.0 reference implementation, and I specify a reproducible evaluation protocol with fault injection, baselines, and outcome measures.

\subsection*{1.3 What this paper is and is not}

This is a methods and framework paper. It defines the contract, the invariant, and the verification primitive precisely enough to be implemented and tested, and a reference implementation already exists and is cited. It specifies, but does not report at production scale, a large-scale empirical study; the worked example is illustrative. The cryptographic building blocks are established; the contribution is their composition into a notarized, fail-closed publication gate for serverless federated writes. I present PVDM as a proposed specification accompanied by a reference implementation, not as a ratified standard: the normative requirements of Section 13 state what conformance would mean, and Section 13.5 outlines the open-governance path that standardization would require. Where this paper uses normative language, it describes the proposed contract, not an already-binding standard.

\subsection*{1.4 A motivating incident (illustrative)}

To make the failure concrete, consider a composite of incidents familiar to practitioners, presented as an illustration rather than a specific event. A payments domain runs a nightly backfill that reads the day's settled transactions from an operational store and writes a curated partition to a shared lakehouse table that finance queries for regulatory reporting. One night an upstream pagination bug causes the reader to stop early on one shard, so roughly two percent of transactions are never read. The write succeeds: every Parquet file it produces is well formed, every object passes its storage checksum, and the table-format commit is atomic and isolated. The orchestrator reports success and the on-call engineer sees a green run.

Nothing in the stack objects, because nothing in the stack is checking the question that matters. The storage layer confirms the bytes it received are intact. The table format confirms the commit applied atomically and readers see a consistent snapshot. A row-count data-quality test, if one runs, compares against an expected range that two percent variance does not exceed. The partition is now durable, atomic, isolated, consumer-visible, and missing two percent of its rows. Finance files a report from it. The error surfaces weeks later during reconciliation, and the cost is not a rerun but a restatement and an audit finding.

This composite is illustrative, but the failure mode it depicts is not hypothetical. Practitioners report the same pattern in production: an analytics model that delivered a few hundred rows where more than eighty thousand were expected, discovered only when business users noticed the discrepancy [25], and the broadly observed hazard that a pipeline can report success while an upstream system silently halves its output or writes zero rows [26]. The recurring shape is exactly the one PVDM targets: the job succeeds, the data is wrong, and the gap between the two is invisible to the success signal the platform reports.

The lesson is specific. Every guarantee in the pipeline was about the transaction or the storage, and each held. The missing guarantee was about the content: that the set of rows published equals the set that should have been published. PVDM exists to add exactly that guarantee, as a mandatory gate, so that the green run and the correct partition are the same event rather than two things an engineer hopes are aligned.

\section*{2. Background and Related Work}

Several established patterns solve adjacent problems, and PVDM composes with them rather than replacing them. I treat each in turn, because the value of PVDM is defined precisely by the gap each leaves open.

\textbf{Data mesh.} The data mesh [12] organizes data ownership around domains that publish data as products. It is an organizational and ownership model; it does not define a write-transaction primitive or a content-integrity gate. PVDM supplies a technical publication contract that a domain in a mesh can implement, and its DataProductContract is the registry-facing envelope.

\textbf{Medallion architecture.} Bronze, silver, and gold layering organizes refinement stages. It does not provide a cryptographic source-to-sink proof between stages. PVDM governs the silver-to-gold publication step with a proof.

\textbf{Outbox and saga.} The outbox pattern [14] makes async side effects reliable and the saga pattern [13] coordinates distributed transactions through compensations. Both reason about message delivery and workflow state, using logs rather than a content-equivalence proof; neither provides a multiset gate over the published data.

\textbf{Two-phase commit.} Two-phase commit [15] gives atomic commit across resources but relies on a central coordinator and does not fit domain-owned serverless execution; it also says nothing about content equivalence.

\textbf{Warehouse tests and job bookmarks.} Row-level warehouse tests assert data-quality rules after load, and job bookmarks track an incremental ETL cursor. Neither is a multiset hash, and neither is a mandatory pre-snapshot gate; tests can pass on sampled rules while rows are silently missing, and a bookmark records executor state, not a source-to-sink proof.

\textbf{Write-Audit-Publish.} WAP, supported by Iceberg branching, stages writes to a branch, audits, then publishes. PVDM is closest to WAP and can be read as a hardening of it. The difference is that WAP audits are typically queries, and publication is possible once they pass; PVDM requires a cryptographic multiset proof, notarized independently, and makes publication impossible without a PASS.

\textbf{Open table format commit protocols.} Apache Iceberg, Delta Lake, and Apache Hudi are the dominant open table formats, and all three give atomic commits and snapshot isolation over object storage. They differ in how they serialize concurrent writers. Iceberg and Delta use optimistic concurrency control, retrying a commit whose snapshot base moved.

Apache Hudi organizes writes on a commit timeline with multi-version concurrency control and, in Hudi 1.0, adds Non-Blocking Concurrency Control [27] so that concurrent writers to the same table proceed without blocking and are ordered by their completion time. These mechanisms make a commit atomic, isolated, and increasingly safely concurrent, but each validates that a commit is well formed and serializable, not that the multiset of rows it publishes equals the multiset the job intended to write. PVDM is orthogonal and complementary: it composes on top of whichever format is in use (its reference gate targets Iceberg) and supplies the content-equivalence proof these protocols leave to the writer's exit status.

Hudi's non-blocking control is directly relevant to the multi-writer operating mode of Section 10.2: it removes a liveness bottleneck for concurrent publishing, yet it does not by itself establish whether a non-blocking concurrent write dropped or duplicated rows, which is exactly what the PVDM gate checks.

\textbf{Recent catalog and format developments.} The surrounding ecosystem is moving quickly, and I situate PVDM against it without claiming integration with each piece. Apache Polaris [28] implements the Iceberg REST Catalog API with role-based access control and short-lived credential vending for multi-engine access, which is a natural host for the Steward-side catalog and the cross-account grants of Section 5. Iceberg v3 [29] adds deletion vectors, row lineage, and a VARIANT type. These change how catalogs enforce access and how deletes and semi-structured values are represented; they do not change the content-integrity gap PVDM addresses, and the reference gate operates against the Iceberg v2 and v3 tables that PyIceberg produces.

\textbf{The end-to-end argument.} The end-to-end argument in system design [1] holds that a correctness function is best placed at the endpoints, because lower layers cannot fully guarantee it. Content integrity for a data consumer is such a function; inferring it from a writer's exit code is the lower-layer placement the argument warns against, and verifying it at publication against an independent statement of intent is the endpoint placement PVDM adopts.

\textbf{Integrity primitives.} Merkle trees [2] authenticate ordered data with a single root and support membership proofs, and object stores use per-object checksums in the same spirit. Merkle roots are order-sensitive: the same rows written in a different order or split across a different number of files produce a different root. Incremental multiset hash functions [3], [4] map a multiset to a digest that is independent of insertion order and can be updated incrementally, which is exactly what is needed to compare the intended and written content of a write that fans out across many serverless tasks and files. PVDM uses a multiset hash as its content proof and reserves per-file checksums for detecting physical corruption within a file.

\textbf{Exactly-once and idempotency.} A large body of work in stream processing and messaging pursues exactly-once semantics through idempotent producers, transactional commits, and deduplication keys [20]. These mechanisms prevent duplicate effects and, in mature systems, are trustworthy. PVDM does not compete with them; where a team already has an exactly-once path it trusts, the pattern's own guidance is to skip the gate. The gap PVDM fills is the common case where the write path is not exactly-once end to end, where a serverless retry can duplicate or a partial read can drop, and where no cryptographic record proves that the final published content is neither short nor doubled. Idempotency suppresses duplicate requests; it does not prove that the committed dataset equals the intended dataset.

\textbf{Authenticated data structures.} Merkle-based authenticated data structures, and their use in verifiable databases and transparency logs [19], provide efficient proofs that a value is included in a committed structure. They are powerful when the question is membership or inclusion against a known root. The question PVDM asks is different and set-oriented: does the multiset of published rows equal the multiset of intended rows, regardless of layout. An inclusion proof against an ordered structure does not answer this directly, which is why the content proof is a multiset digest rather than a Merkle root, with per-file Merkle-style checksums retained only for the orthogonal purpose of physical tamper evidence.

\subsection*{2.1 Where PVDM sits}

\vspace{8pt}\noindent\small\textit{Table 2. PVDM names and structures}\vspace{4pt}

\begin{longtable}{|p{0.30\textwidth}|p{0.30\textwidth}|p{0.30\textwidth}|}
\hline
\textbf{\textbf{Pattern}} & \textbf{\textbf{What it solves}} & \textbf{\textbf{What it does not provide}} \\
\hline
Data mesh & Decentralized ownership, data as product & A write-transaction primitive \\
\hline
Medallion & Bronze/silver/gold layering & A cryptographic source-sink proof \\
\hline
Outbox & Reliable async side effects & A multiset-equivalence gate \\
\hline
Saga & Distributed compensations & A proof notary; logs, not math \\
\hline
Two-phase commit & Atomic commit across resources & Domain-owned serverless fit; content proof \\
\hline
Warehouse tests & Row-level quality rules & Multiset hash; mandatory pre-snapshot gate \\
\hline
Job bookmarks & Incremental ETL cursor & Source-sink content proof \\
\hline
Write-Audit-Publish & Staged, audited publication & A mandatory cryptographic proof gate \\
\hline
Table formats (Iceberg, Delta, Hudi) & Atomic commit, snapshot isolation, writer concurrency (OCC; Hudi NBCC) & Content-equivalence proof: written multiset equals intended \\
\hline
\textbf{PVDM} & \textbf{Proof-gated publication: written multiset equals intended, notarized, fail-closed} & \textbf{Not a storage format; reuses format atomicity/isolation and object durability} \\
\hline
\end{longtable}

\subsection*{2.2 Why the composition is not obvious}

The primitives PVDM uses are individually well known, so the contribution must be defended as a composition that resolves conflicts the pieces create when combined in a serverless federated setting. Several natural designs fail for specific reasons, and ruling them out is what fixes the shape of the protocol.

\begin{itemize}
\item \textbf{A Merkle root over the written files does not work.} A Merkle tree is order-sensitive, so the same logical rows written in a different file order or partition layout, which is normal for parallel serverless writers, produce a different root and a false mismatch. The comparison must be over a multiset, which is what forces an order-independent, incremental hash rather than the more familiar tree.

\item \textbf{Per-file checksums do not work.} They prove each file is individually intact but say nothing about whether the set of rows equals the intended set; a dropped or duplicated file passes every per-file check. The gate has to bind a digest of the whole multiset to a declared intent, not to the bytes that happened to land.

\item \textbf{Verifying inside the producer does not work.} If the party that writes the data is also the party that computes and stores the proof, a faulty or compromised producer certifies its own corruption and can delete the evidence. This is what forces the proof and its keys into a separate Steward account, and it is why the keyed hash (not a plain hash) is load-bearing: the producer must not be able to forge a passing digest.

\item \textbf{A single execution model does not work.} Serverless functions are time-bounded and retried, so a long backfill cannot be one atomic execution, yet naive retries duplicate or drop chunks. Separating a container clock from a workload clock, and replaying completed chunks rather than rewriting them, is the non-obvious step that makes exactly-once progress compose with the fail-closed gate.

\item \textbf{Verifying a transforming stage against itself does not work.} For aggregation or join stages there is no pre-existing intent to compare against, and comparing the producer's output to the producer's own recomputation proves only that the code is deterministic, not that it is correct. This is why PVDM has to distinguish conformance profiles and require an independent oracle for transforming stages rather than pretending one mechanism covers every stage.
\end{itemize}

Each exclusion removes a simpler design, and what remains is the specific arrangement PVDM specifies. The engineering claim is not that any one primitive is new, but that this is the arrangement in which they actually deliver a fail-closed content guarantee across a federated, serverless boundary.

\subsection*{2.3 Notation and conventions}

The following symbols are used throughout. Requirement labels (N1 to N20) and profile labels (A, T, O) are defined in Section 13.

\begin{table}[H]
\centering\small
\begin{tabular}{|p{0.45\textwidth}|p{0.45\textwidth}|}
\hline
\textbf{\textbf{Symbol}} & \textbf{\textbf{Meaning}} \\
\hline
M, W & the intended and the written multiset of rows for a chunk \\
\hline
c(r) & canonicalization of row r to a deterministic byte string \\
\hline
k & the Steward-held secret key for the keyed hash \\
\hline
$h_k$ & keyed hash (HMAC-SHA256) modeled as a pseudorandom function \\
\hline
$D_k(M)$ & multiset digest of $M$; $D_k^{id}$ and $D_k^{ct}$ are the identity and content projections \\
\hline
VRP & Verifiable Reconciliation Proof, the PASS/FAIL comparison of intent and written digests \\
\hline
$C, W$ & a chunk $C$ within a workload $W$ \\
\hline
\end{tabular}
\vspace{8pt}
\caption*{Table 3. Notation and symbols used throughout the paper.}
\end{table}

Requirement levels use RFC 2119 [8] keywords (MUST, SHOULD, MAY). Numerical values in worked examples are illustrative and are labeled as such.

\section*{3. Threat Model and the PVDM invariant}

\textbf{Actors.} An untrusted Producer (domain-owned serverless compute that may fail partially, retry, or emit wrong content); an independent, trusted Steward notary (holds the proofs and checkpoints, performs the proof-gated catalog commit, and lives in an account the producer cannot delete from); a trusted Publisher zone (the consumer-facing lakehouse storage and catalog); and a trusted object store and table catalog that provide durability, atomicity, and isolation as specified by the format.

\textbf{Defended faults.} Silent content faults that would otherwise become a valid commit: dropped rows or partitions, truncated files, duplicated chunks (including from at-least-once serverless retries), and value corruption, plus physical corruption of staged files detected by per-file checksums.

\textbf{Out of scope.} A compromised Steward or catalog; a wrong statement of intent (if the intended-content proof is derived from already-corrupt source data, PVDM will faithfully publish content matching that wrong intent, a garbage-in property, not a gate defect); and confidentiality (PVDM is an integrity mechanism and composes with, but does not provide, encryption).

\textbf{Trust assumptions (consolidated).} The guarantees below rest on exactly these assumptions, stated once here so no later claim depends on an unstated premise:

\begin{itemize}
\item \textbf{A1 Steward key confidentiality.} The multiset-hash key and the proof-signing key are held only by the Steward, are managed in a key-management service, and are never exposed to the Producer. This is what makes the digest unforgeable to the party being checked (Section 7.3).

\item \textbf{A2 Steward and catalog integrity.} The Steward notary and the table catalog execute their specified logic and are not compromised. A compromised Steward or catalog is out of scope (above).

\item \textbf{A3 Account independence.} The Steward account is administratively independent of the Producer, so the Producer cannot delete or alter proofs, checkpoints, or the audit record of its own failures.

\item \textbf{A4 Consumer read path.} Consumers observe data only through the gated catalog snapshot; staged or branch data is not reachable by a consumer read. Access-path integrity is required for the safety invariant to mean what it says (Section 12.14).

\item \textbf{A5 Format primitives.} The object store and table format provide the durability, atomicity, and snapshot isolation they specify; PVDM adds content fidelity on top and does not re-implement them.
\end{itemize}

\textbf{The PVDM invariant.} For every chunk C in a workload W:

commit\_metadata(C)  =>  VRP(C) = PASS
VRP(C) = FAIL       =>  not exists snapshot' : consumers\_visible(snapshot')

In words: metadata is committed for a chunk only if its verification proof passes, and if the proof fails no consumer-visible snapshot referencing that chunk ever exists. The executor exit code is irrelevant to the decision.

\textbf{Safety (fail-closed).} Every consumer-visible snapshot produced under PVDM has a passing content proof. Equivalently, a content fault that changes the written multiset relative to intent cannot reach a consumer, because the catalog commit that would make it visible is gated on the proof and the proof is computed and held by an independent notary.

\textbf{Liveness (no false block, bounded overhead).} If the written multiset equals the intended multiset, the proof passes and the snapshot publishes, with overhead bounded by the cost of computing the two multiset digests once plus one catalog commit. PVDM trades publication latency for the verification step; Section 10 measures that overhead, and the pattern is explicitly batch and backfill oriented rather than sub-second.

\subsection*{3.1 Formal guarantees}

I state the guarantees the gate targets and sketch why they hold under the threat model. The digest $D_k$ referenced below is defined formally in Section 7.1, and all symbols are collected in the notation table of Section 2.3. The proofs are informal and reduce to the digest security of Section 7.3 and standard signature and ledger arguments.

On mechanizability: the safety and tamper-evidence properties are of the invariant-over-a-state-machine form that a specification language such as TLA+ expresses naturally, and the digest-security reduction is of the game-based form suited to a proof assistant such as Coq or an EasyCrypt development. The obstacle to full mechanization is not a known gap in the arguments but the engineering effort of modeling the catalog commit and the cryptographic assumptions faithfully; I identify it as a separate formal-methods effort (Section 18) rather than claim a machine-checked result here.

\textbf{Theorem 1 (fail-closed content safety).} Under the assumptions that HMAC-SHA256 is a pseudorandom function and the Steward key is confidential, every consumer-visible snapshot produced by PVDM has a written content multiset whose digest equals the intended digest recorded in a valid, signed, PASS proof. Sketch: the Metadata gate commits only when it verifies a Steward signature over a proof whose verdict is PASS and whose per-file digests equal the digests of the bytes being published. A PASS verdict requires $D_k(intent)$ = $D_k(written)$ by construction.

To publish content that diverges from intent, an adversary must either produce a divergent written multiset with a colliding digest, which contradicts digest security, or forge or alter the signed proof, which contradicts signature unforgeability. Hence no divergent content is published, which is the fail-closed property.

\textbf{Theorem 2 (tamper evidence at commit).} If the bytes presented at commit differ from the bytes verified, the commit is rejected. Sketch: the proof binds a per-file content digest for every file, and the gate recomputes those digests over the exact bytes being published and requires equality. Any change to any published file changes at least one digest, so the equality check fails and the commit is rejected. This closes the time-of-check to time-of-use window between Verify and Metadata.

\textbf{Theorem 3 (replay and misdirection freedom).} A valid proof cannot be used to publish a chunk, target, or workload other than the one it was issued for, nor can it be used more than once. Sketch: the signed proof body includes the workload identifier, chunk identifier, target, partition, and a fresh nonce. The gate checks that the target matches the commit target and that the nonce has not been seen, and any change to these fields invalidates the signature. Therefore a proof is bound to a single commit context and a single use.

\textbf{Scope of the proofs.} These theorems are conditional on the trust assumptions A1 to A5 of Section 3, and they are informal sketches, not mechanized proofs; I deliberately do not claim more than the informal arguments support, and a machine-checked proof of Theorem 1 is stated as future work. Their boundaries form part of the precise statement of the protocol's guarantees. They do not guarantee fidelity to a true upstream state when the intent is derived from a corrupt source read (assumption breaks at the source, Section 12.1); they do not hold if the Steward key is compromised or the notary colludes (a direct violation of A1 and A2, Section 12.6); and they concern content integrity, not confidentiality or availability, the latter being consciously traded away by the fail-closed choice (Section 12.9). In particular, the fail-closed safety claim is a claim about an uncompromised gate operating under A1 to A5, and I make no stronger claim; a compromised Steward voids the guarantee and is out of scope by construction, not covered by these proofs.

\section*{4. The PVDM Protocol}

PVDM has four mandatory phases. The acronym is exactly four letters; optional business-rule or integrity checks may run before the Physical phase but are not a fifth phase.

\begin{figure}[H]
\centering
\includegraphics[width=0.9\textwidth,keepaspectratio]{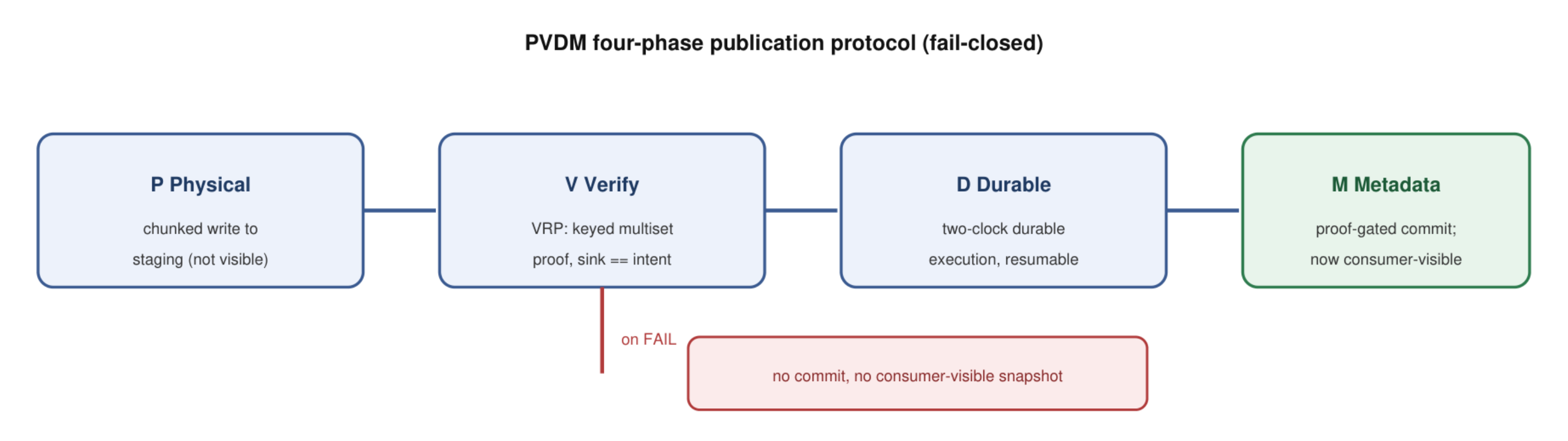}
\vspace{8pt}
\caption*{Figure 1. The four-phase publication protocol. Data written in the Physical phase stays invisible to consumers until the Verify phase produces a passing proof and the Metadata phase commits; a failed proof comm}
\end{figure}

\textbf{P: Physical.} Domain-owned serverless compute writes data to storage in a form that is not yet consumer-visible. Uncommitted files can be rolled back. The invariant of this phase is rollbackability: if the invocation is interrupted, no orphan data becomes visible, and staged checkpoints record progress. A violation of this phase shows up as orphan Parquet files in the lakehouse.

\textbf{V: Verify.} A content proof is computed over the staged data using an order-independent multiset hash (Section 7), and compared against a proof of the intended data. The proof and its result are stored by the Steward notary, not by the producer. The invariant is that a passing multiset proof is required to proceed; its violation symptom is silent row loss reaching consumers. This phase is the gate: validate\_then\_commit.

\textbf{D: Durable.} For workloads that span more than one serverless invocation, completed chunks are replayed, not rewritten, using durable orchestration keyed by a workload identifier. The invariant is exactly-once progress across retries; its violation symptom is duplicate data on a function retry.

\textbf{M: Metadata.} The table catalog commit is performed last and only if the proof passed. The invariant is that the catalog commit is the final, proof-gated step; its violation symptom is phantom snapshots that reference unverified data.

The control flow is Physical then Verify; on a passing proof, Durable then Metadata; on a failing proof, the protocol stops and no snapshot is created, so consumers are unchanged.

\begin{table}[H]
\centering\small
\begin{tabular}{|p{0.30\textwidth}|p{0.30\textwidth}|p{0.30\textwidth}|}
\hline
\textbf{\textbf{Phase}} & \textbf{\textbf{Invariant}} & \textbf{\textbf{Violation symptom}} \\
\hline
Physical & Uncommitted files can be rolled back & Orphan Parquet in lakehouse \\
\hline
Verify & Multiset proof PASS required & Silent row loss reaches consumers \\
\hline
Durable & Completed chunks replay, not rewrite & Duplicate data on retry \\
\hline
Metadata & Catalog commit is last, proof-gated & Phantom snapshots \\
\hline
\end{tabular}
\vspace{8pt}
\caption*{Table 4. PVDM protocol failure state transitions.}
\end{table}

\subsection*{4.1 The protocol in pseudocode}

The following pseudocode makes the phase ordering and the gate explicit for a single chunk. The Producer executes Physical; the Steward executes Verify and the proof-gated Metadata commit; Durable wraps the whole sequence so that a resumed invocation re-enters at the last completed step rather than the beginning.

\begin{figure}[H]
\centering
\includegraphics[width=0.9\textwidth,keepaspectratio]{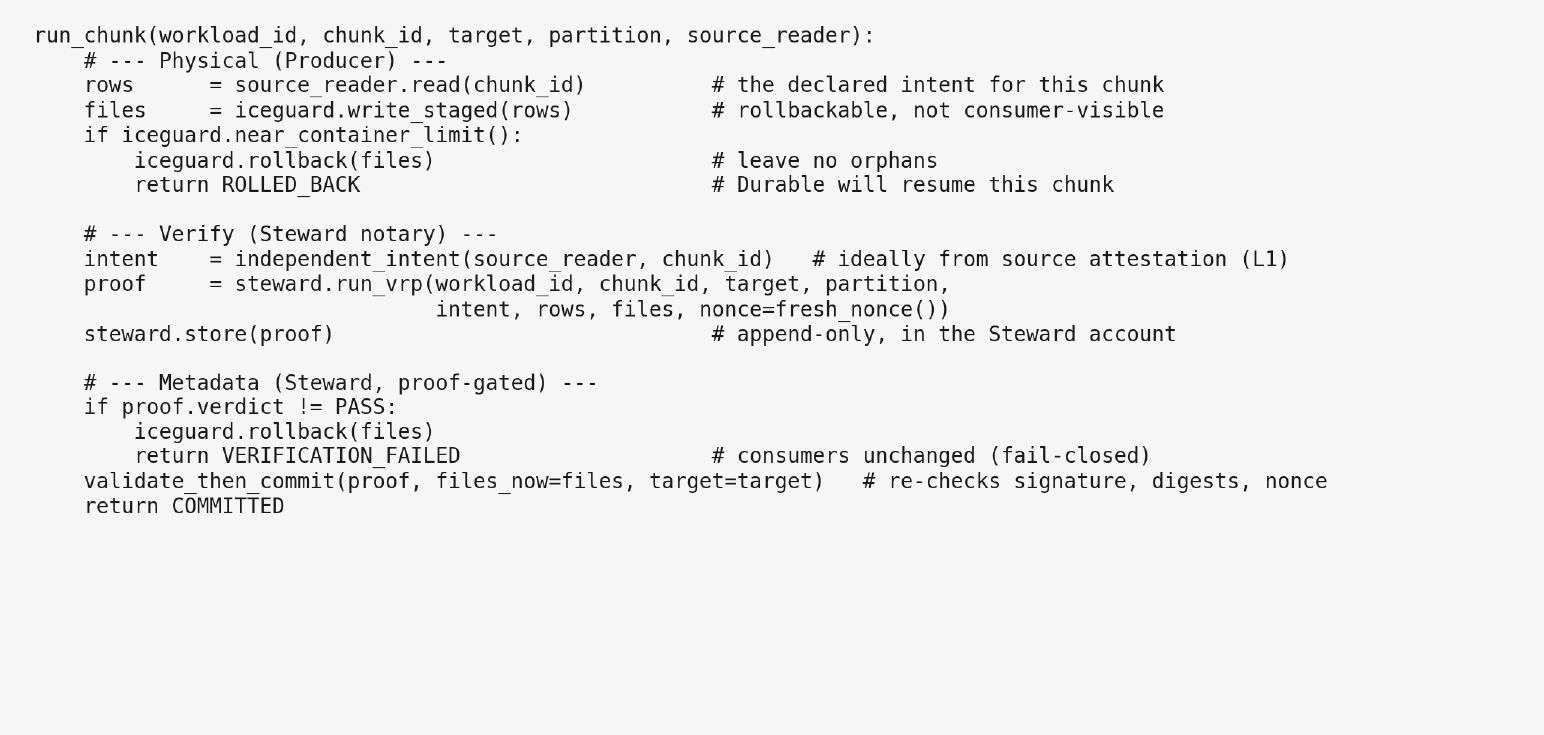}
\vspace{8pt}
\caption*{Figure 2. Listing 1. Per-chunk execution: Physical, Verify, and proof-gated Metadata (Section 4.1).}
\end{figure}

\textit{Runnable reference source: \url{https://github.com/vaquarkhan/Proof-gated-publication-PVDM}, validation/pvdm\_gate.py (run\_vrp, validate\_then\_commit).}

The Durable phase is not a separate branch in this listing; it is the execution context in which run\_chunk runs. A resumed segment reconstructs which chunks already reached COMMITTED from the durable step log and the Steward checkpoints, and it re-executes only chunks whose last recorded outcome was ROLLED\_BACK, which is what gives the Durable invariant that completed chunks replay rather than rewrite.

\subsection*{4.2 Failure state machine}

Each chunk moves through a small state machine whose terminal states are the three outcomes a coordinator can return. From STAGING a chunk goes to VERIFYING on a complete write or to ROLLED\_BACK if the container clock nears its limit. From VERIFYING it goes to COMMITTING on a PASS or to VERIFICATION\_FAILED on a FAIL.

From COMMITTING it goes to COMMITTED if validate\_then\_commit accepts, or back to VERIFICATION\_FAILED if the commit gate rejects (for example a post-verify tamper). ROLLED\_BACK is not terminal for the workload: the workload clock resumes the chunk in a new segment, and only when every chunk reaches COMMITTED does the workload succeed. VERIFICATION\_FAILED is terminal for the attempt and routes to an operator-visible repair path that re-reads and re-verifies rather than blindly retrying, so a genuine data fault is surfaced instead of being masked by a retry.

This state machine is what connects the per-chunk invariant of Section 3 to the per-workload behavior an operator observes.

\begin{figure}[H]
\centering
\includegraphics[width=0.9\textwidth,keepaspectratio]{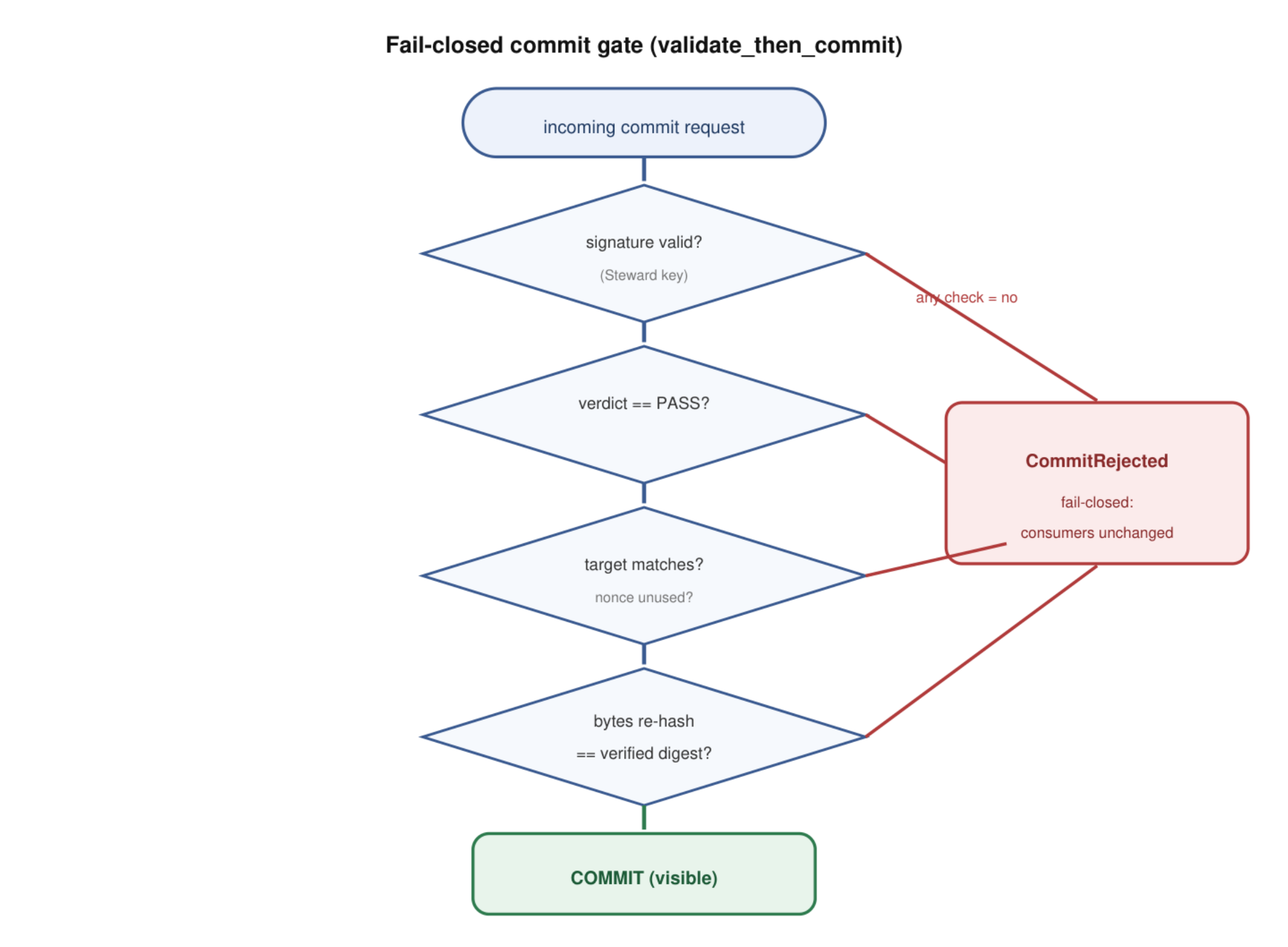}
\vspace{8pt}
\caption*{Figure 3. The proof-gated commit gate. Signature validity, a PASS verdict, target and nonce checks, and a re-hash of the exact bytes being published must all hold, or the commit is rejected and consumers stay u}
\end{figure}

\section*{5. The Three-Account Federated Model}

PVDM separates duties across three accounts so that the entity that produces data is not the entity that attests to its correctness.

\begin{figure}[H]
\centering
\includegraphics[width=0.9\textwidth,keepaspectratio]{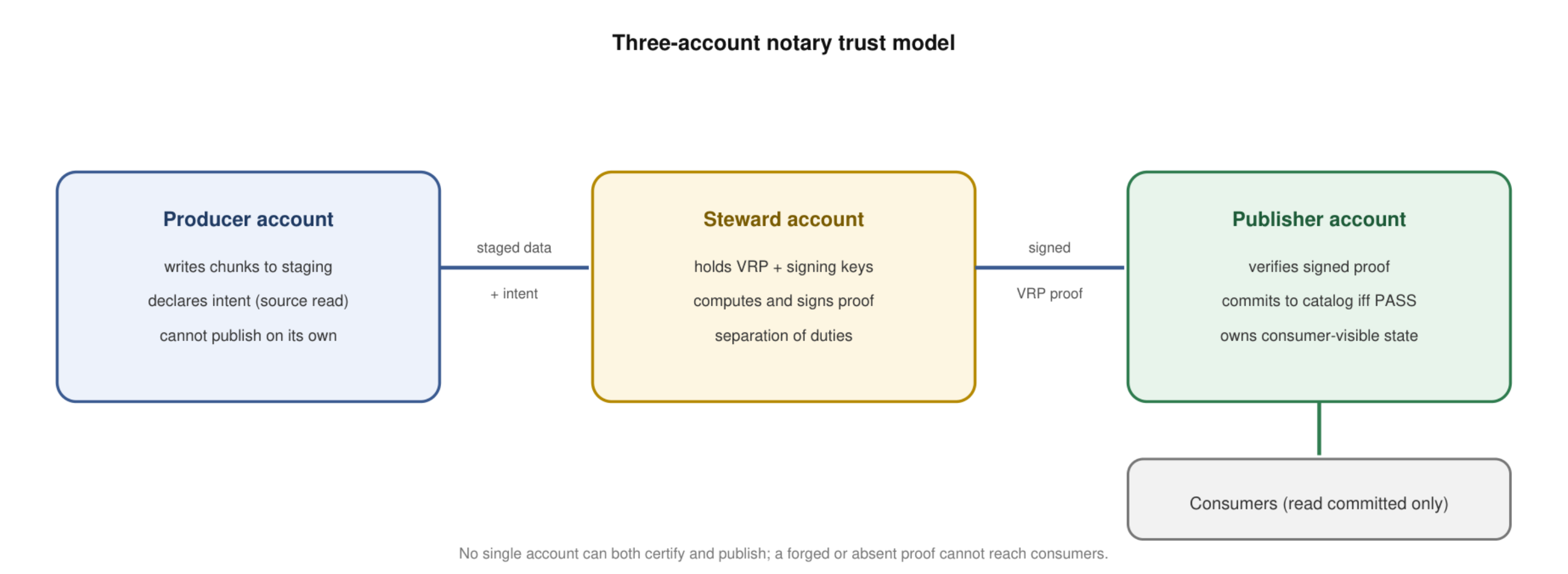}
\vspace{8pt}
\caption*{Figure 4. The three-account trust model. The Producer stages data and declares intent, the Steward holds the keys and signs the proof, and the Publisher commits to the consumer-visible catalog only on a passing}
\end{figure}

\begin{itemize}
\item \textbf{Producer.} Domain autonomy. Holds the serverless functions, the durable orchestration, and the domain code that performs the Physical write.

\item \textbf{Steward.} Federated governance and notarization. Holds the verification proofs, the checkpoints, and the table catalog, and performs the proof-gated Metadata commit. Because proofs and checkpoints live in the Steward account, the producing domain cannot unilaterally delete the audit evidence.

\item \textbf{Publisher.} Blast-radius isolation. Holds the consumer-facing lakehouse storage and the reader-visible table, so a producer fault cannot directly damage the consumer zone.
\end{itemize}

The Steward is the notary. Placing proofs and checkpoints where the domain cannot delete them is what makes the audit trail trustworthy in a federated setting, and it is the structural reason PVDM is stronger than a producer-local quality check. Storing proofs in the Producer account is an explicit anti-pattern, because it lets a domain erase the record of its own failures.

\section*{6. Segmented Serverless Execution: Two Clocks}

Serverless functions are time-bounded and retried, so a long backfill cannot run as a single execution. PVDM models this with two clocks linked by a workload identifier.

\begin{itemize}
\item \textbf{Container clock.} Owned by an in-process watchdog (IceGuard) within a single function invocation, bounded by the platform limit (for example 900 seconds on AWS Lambda). As the invocation approaches its limit, the watchdog rolls back uncommitted work and reports that the segment rolled back cleanly.

\item \textbf{Workload clock.} Owned by the durable orchestration layer (a durable execution SDK plus a state machine), bounded by the full backfill budget (for example 5400 seconds). It resumes the workload in a new segment and replays only completed steps.
\end{itemize}

The linkage key is a workload\_id that ties checkpoints, proofs, and durable replay across segments. The sequence is: the orchestrator invokes a segment with a workload\_id; the segment writes chunks under the watchdog; if the container clock nears its limit, the watchdog rolls back uncommitted work and the segment returns a rolled-back status; the orchestrator resumes with the next segment, which replays completed steps only rather than rewriting them; when a chunk completes it is committed. This is what gives the Durable phase its exactly-once-progress property across the natural retries of serverless compute.

The resumption logic that the other three phases delegate to Durable is made explicit below. Unlike run\_chunk in Section 4.1, which executes one chunk, resume\_workload is the durable driver: it is idempotent under re-invocation because it reads the committed outcome of each chunk from the Steward step log before deciding whether to act, so a segment that is retried after a crash neither rewrites a COMMITTED chunk nor loses a ROLLED\_BACK one.

\begin{figure}[H]
\centering
\includegraphics[width=0.9\textwidth,keepaspectratio]{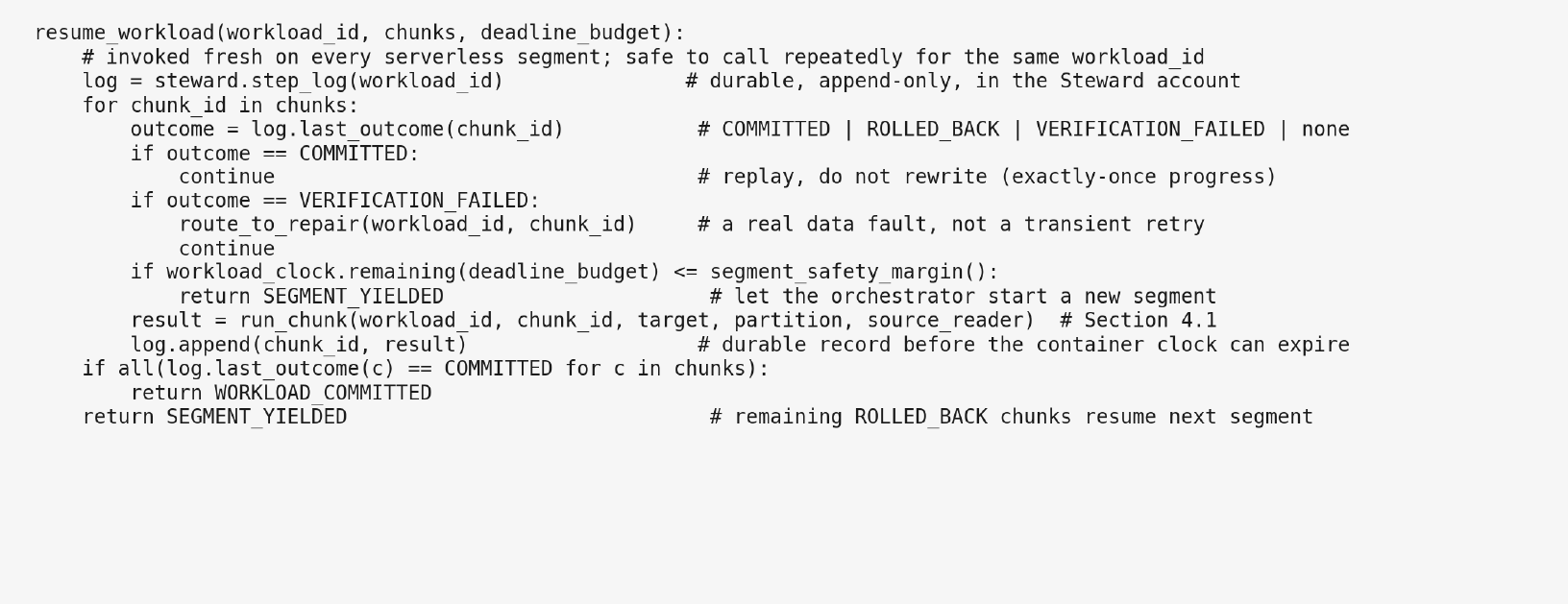}
\vspace{8pt}
\caption*{Figure 5. Listing 2. Durable resume across serverless segments (exactly-once progress).}
\end{figure}

\textit{Runnable reference source: \url{https://github.com/vaquarkhan/Proof-gated-publication-PVDM}, validation/pvdm\_gate.py.}

Two invariants make this correct. First, the step log is written in the Steward account and consulted before any action, so the completed/incomplete decision survives a container-clock expiry mid-segment. Second, run\_chunk is itself fail-closed, so a chunk that yields between Physical and Metadata leaves no consumer-visible snapshot and is simply re-executed; only a COMMITTED record suppresses re-execution.

\section*{7. The Verification Primitive: An Order-Independent Multiset Proof}

The Verify phase compares what was written to what was intended using a Verifiable Reconciliation Proof (VRP). The proof computes multiset hashes separately over the declared identity fields and the content fields of the rows, so it detects not only missing or extra rows but also in-place value mutation and schema drift, and it compares source rows against the sink Parquet. The comparison must be invariant to the things a distributed serverless write legitimately varies: the number of files, the partitioning, and the order of rows within and across files. Two runs that persist the same rows in a different physical layout are content-equivalent and must both pass, and a run that drops or duplicates even a single row must fail.

An order-sensitive digest such as a plain Merkle root over a file or a row sequence cannot express this, because it changes when the layout changes even though the content is identical. The correct primitive is an incremental multiset hash: a function that maps a multiset of rows to a digest such that the digest is independent of the order in which rows are added and can be updated incrementally as each task contributes its rows. The Verify phase computes the multiset digest of the intended rows and the multiset digest of the written rows and requires them to be equal. Because the digest is a multiset hash rather than a set hash, a duplicated row changes the digest and is caught, not silently absorbed.

Per-file Merkle-style checksums remain useful and are complementary: they detect physical corruption of an individual staged file (bit-rot, truncation) before the multiset comparison, and they localize a fault to a file. The multiset proof detects logical content divergence between intent and result; the per-file checksum detects physical damage to a stored file. PVDM uses both, at different granularities.

The proof is what an auditor later re-checks offline: given the stored intended digest and the stored written digest for a chunk, an independent verifier can confirm that the published content matched intent, without re-running the pipeline, which is the offline-verifiable evidence the Steward notary exists to provide.

\subsection*{7.1 Formal construction of the VRP digest}

Let $R$ be the universe of rows and let $c: R \to \{0,1\}^*$ be a canonicalization function that maps a row to a deterministic byte string (sorted fields, normalized types, explicit nulls, fixed decimal scale, UTC timestamps). Let $k$ be a secret key held by the Steward and let $h_k: \{0,1\}^* \to \{0,1\}^{256}$ be a keyed hash (HMAC-SHA256 [5], [6], [7]) that I model as a pseudorandom function. For a multiset $M$ of rows I define the digest

{\small
\begin{equation}
D_k(M) = \left( \sum_{r \in M} \mathrm{int}(h_k(c(r))) \right) \bmod 2^{256}
\end{equation}
}

where $\mathrm{int}()$ interprets the 256-bit hash as an integer and the sum is taken with multiplicity, so a row that appears $t$ times contributes $t$ times. Two projections are used: the identity digest $D_k$ restricted to the declared identity fields, written $D_k^{id}$, and the content digest over the full canonical row, written $D_k^{ct}$. A chunk verifies when both the identity and content digests of the intent multiset equal those of the written multiset:

{\small
\begin{align*}
&\mathrm{VRP}(\text{int}, \text{wrt}) = \text{PASS} \iff \\
&\quad D_k^{id}(\text{int}) = D_k^{id}(\text{wrt}) \\
&\quad \wedge\; D_k^{ct}(\text{int}) = D_k^{ct}(\text{wrt})
\end{align*}
}

The identity projection localizes faults: if $D^{id}$ matches but $D^{ct}$ does not, some row kept its identity but changed a value, which is the signature of an in-place mutation; if $D^{id}$ diverges, rows were added or dropped. This two-projection form mirrors the veridata-recon design and gives operators a cheap first classification of a failure without a full row-level diff.

\subsection*{7.2 Properties of the digest}

The construction has four properties that the pattern relies on, each following directly from the definition. First, order independence: modular addition is commutative and associative, so $D_k(M)$ does not depend on the order in which rows are enumerated, hence it is invariant to file order and to the order of rows within a file. Second, layout independence: because $D_k$ is defined over the multiset of canonical rows and not over files, splitting the same rows across a different number of files or partitions yields the same digest, so a legal repartition does not change the proof.

Third, multiplicity sensitivity: because the sum is taken with multiplicity, changing how many times any row appears changes the digest, so duplicates and drops are both detected. Fourth, incrementality: $D_k(M1 union M2)$ = ($D_k(M1)$ + $D_k(M2)$) mod $2^{256}$ for disjoint enumeration, so the digest can be accumulated as per-shard partials and combined without materializing the whole multiset, which is what makes the scheme practical at scale.

A necessary clarification about where hashing runs, because it interacts with key isolation. The multiset hash is keyed with a Steward-held secret that, by requirement N2, the Producer must not read. Therefore the keyed hashing is a Steward-trust-domain operation: verification workers under the Steward compute the partial digests over the source (for intent) and over the sink files (for written content), and the Steward combines them.

The Producer performs the Physical write but does not hold the hashing key and does not produce the authoritative digest. This preserves both properties at once: the digest is parallel and incremental across many workers, and the key never leaves the Steward domain. Partial digests are therefore produced inside the trust boundary that holds the key, so a Producer task cannot forge a partial, and the combination step additionally requires all partials to share a key epoch (an enforced check) before they are summed.

\subsection*{7.3 Security argument}

I argue informally that, without the key k, a computationally bounded adversary cannot construct a written multiset W different from the intended multiset M with $D_k(W) = D_k(M)$. Because $h_k$ is modeled as a pseudorandom function, each element contribution $\mathrm{int}(h_k(c(r)))$ is, to an adversary without k, indistinguishable from a uniformly random 256-bit value. Finding a nonidentical multiset with the same modular sum is then equivalent to finding a nontrivial subset-sum collision over pseudorandom 256-bit values, which is infeasible for a bounded adversary by the same reasoning that establishes the security of incremental multiset hash functions in the keyed setting.

The key is essential: with an unkeyed or public element hash the contributions are known, and the adversary can solve a linear equation to substitute rows, which is exactly the collision the validation suite constructs and the keyed digest defeats. The security therefore reduces to two assumptions I state plainly: that HMAC-SHA256 behaves as a pseudorandom function, and that the key k is confidential to the Steward and never exposed to the Producer.

\subsection*{7.4 Worked digest example (illustrative)}

Consider three rows with identity field payment\_id and content field amount. Canonicalization yields three byte strings; the Steward computes $h_k$ over each, obtains three 256-bit integers, and sums them mod $2^{256}$ to get $D_k^{ct}$. If the writer drops the second row, the written sum is missing that row's contribution, so the written $D_k^{ct}$ differs from the intended $D_k^{ct}$ by exactly that contribution and VRP fails.

If instead the writer changes the second row's amount, its canonical bytes change, its element hash changes to an unrelated 256-bit value, and again the content digest diverges while the identity digest, computed only over payment\_id, is unchanged, classifying the fault as a mutation. These are the mechanics the validation suite checks; the numbers themselves are not reproduced here because they carry no additional information beyond the divergence.

\subsection*{7.5 Alternative constructions and performance headroom}

The protocol depends on the digest being a keyed, order-independent, multiplicity-sensitive, incremental multiset hash; it does not depend on the specific construction, which is a pluggable choice. I use a keyed sum of HMAC-SHA256 per-row hashes modulo $2^{256}$ (an MSet-Add-Hash in the family of Clarke et al. [3] and Bellare and Micciancio [4]) because it is simple, keyed, and implementable from any standard library, which suits a dependency-free reference gate. It is not the fastest option, and I do not claim it is.

The multiset-hash literature offers constructions with different tradeoffs: MSet-XOR-Hash is fast but only set-collision resistant, MSet-Mu-Hash is multiset-collision resistant, and the Elliptic Curve Multiset Hash (ECMH) [22] achieves a near-optimal digest size and reports throughput on the order of millions of elements per second in optimized native code, well above the reference gate. An order-sensitive Merkle tree is unsuitable for the reason given in Section 2.2, but flat multiset hashing is exactly the shape this comparison needs.

The practical consequence is that the throughput reported in Section 10 (roughly 105,000 to 116,000 rows per second) is a property of a single-threaded standard-library implementation with per-row JSON canonicalization, not a property of the protocol. Because the digest is additive, verification is embarrassingly parallel across shards and rows, and the per-row hash can be replaced with a vectorized or elliptic-curve construction without changing any correctness argument in this paper, since those arguments rest only on the multiset-hash properties of Section 7.2 and the keying assumption of Section 7.3. I measure this headroom directly in Section 10.7: on the same instance, changing only the hash and canonicalization gives a 3.86x speedup and a naive four-core parallelization reaches 4.81x, all while preserving the multiset-hash properties. A production deployment that needs still higher throughput should adopt an optimized or elliptic-curve construction; measuring that at production scale is part of the future work of Section 18.

\subsection*{7.6 Collision resistance: why a corrupted write cannot silently pass}

The gate is only meaningful if a corrupted write cannot produce the same digest as the intended write. I do not prove this from scratch, because I do not need to: the construction here is the keyed MSet-Add-Hash of Clarke et al. [3], a per-element keyed hash summed over the integers modulo a large modulus (here $2^{256}$, which is a ring, not a field). Clarke et al. prove that this construction is multiset-collision resistant under the assumption that the keyed per-element hash behaves as a pseudorandom function (equivalently, in the random-oracle model), and the construction is in the incremental-hashing lineage of the AdHash of Bellare and Micciancio [4].

I therefore state the guarantee of this section as a reduction to those published, peer-reviewed results rather than as a theorem original to this paper: under the pseudorandom-function assumption for HMAC-SHA256, and with the per-element key held by the independent Steward and unavailable to the adversary, finding two distinct multisets with the same digest is computationally infeasible, and the probability that a single dropped, duplicated, or mutated row leaves the additive sum unchanged is on the order of $2^{-256}$. By the birthday bound, roughly $2^{128}$ distinct digests would have to be published before an accidental collision became likely, which is far outside anything physically reachable.

Two caveats I state plainly. First, additive multiset hashes are subject to generalized-birthday (Wagner-style) attacks when the modulus is small or the element hash is unkeyed; the 256-bit modulus and the secret key are exactly what place this construction in the secure, keyed regime that [3] analyzes, and are why requirement N1 mandates a keyed 256-bit hash. Second, the figures above are the security level implied by that reduction, not an unconditional bound: the property is only as strong as the pseudorandom-function assumption on HMAC-SHA256. An independent cryptographer review is welcome, but the claim rests on established results, not on a proof I invented here.

The specific worry an operator raises is whether a dropped, duplicated, or mutated row can leave the digest unchanged. Because the digest is a sum, a single-row fault shifts it by exactly that row's element hash (a drop subtracts it, a duplicate adds it again, a mutation replaces it with an unrelated value), so the digest is unchanged only if that shift is exactly zero modulo $2^{256}$, which for a keyed pseudorandom hash occurs with probability $2^{-256}$. Duplicates in particular cannot be absorbed, which is precisely why the digest is an additive multiset hash and not a set hash: multiplicity is carried in the sum.

I validate this empirically with a dedicated analysis (validation/collision\_analysis.py). First, across two thousand random base multisets I applied every single-row drop, duplicate, and mutation, six thousand faults in all, and every one changed the digest, with zero escapes. Second, because a 256-bit collision cannot be observed directly, I truncated the per-element hash to sixteen, twenty, and twenty-four bits, hashed tens of thousands of distinct elements at each width, and confirmed that the observed collision counts match the exact birthday prediction (for example 2712 observed against 2764 expected at sixteen bits, and 188 against 190 at twenty-four bits).

This confirms that the per-element hash is uniform, which is the assumption under which the 256-bit bound holds, so the extrapolation to $2^{-256}$ is grounded rather than asserted. It also quantifies the fast benchmark hash of Sections 10.4 and 10.8: at thirty-one bits the per-fault escape probability is about $2^{-31}$ and the birthday point is near fifty-four thousand distinct digests, which is adequate to detect the specific injected faults in a benchmark but is exactly why requirement N1 mandates a keyed 256-bit hash for production, where silent collision must be impossible in practice rather than merely unlikely.

\section*{8. Reference Implementation}

A reference implementation exists on AWS in the open-source \href{https://github.com/vaquarkhan/aws-serverless-datamesh-framework}{aws-serverless-datamesh-framework} repository, released under Apache-2.0. It realizes each PVDM phase with named components:

\begin{itemize}
\item \textbf{Coordinator:} IceGuardDurableCoordinator, which enforces the four phases inside a serverless segment.

\item \textbf{Proof gate:} validate\_then\_commit, the Verify-to-Metadata boundary that refuses to commit on a failing proof.

\item \textbf{Proof generator:} VRPProofGenerator (the veridata-recon verification component) that computes the multiset proofs stored by the Steward.

\item \textbf{Contracts:} DomainTransactionBoundary (the declared write boundary) and DataProductContract (the registry-facing product envelope).

\item \textbf{Catalog:} a Glue and Iceberg REST adapter that performs the proof-gated Metadata commit.

\item \textbf{Benchmark and demo:} an evaluation harness (eval/validate\_then\_commit\_benchmark.py) and a local demonstration that a clean write commits while a corrupt write is blocked, runnable without cloud infrastructure.
\end{itemize}

The implementation composes four independently published building blocks, each mapped to a PVDM phase:

\begin{table}[H]
\centering\small
\begin{tabular}{|p{0.30\textwidth}|p{0.30\textwidth}|p{0.30\textwidth}|}
\hline
\textbf{\textbf{Building block}} & \textbf{\textbf{Distribution}} & \textbf{\textbf{Role in PVDM}} \\
\hline
IceGuard & PyPI iceguard & Physical: chunked Parquet writes, timeout rollback, S3 checkpoint/resume \\
\hline
veridata-recon & PyPI veridata-recon & Verify: the VRP (multiset hashes over identity and content fields) per chunk \\
\hline
AWS Durable Execution SDK & PyPI aws-durable-execution-sdk-python & Durable: replay completed steps across 15-minute Lambda segments \\
\hline
PyIceberg Glue REST & PyPI pyiceberg via GlueCatalogConnector & Metadata: SigV4 catalog commit, gated on VRP PASS \\
\hline
\end{tabular}
\vspace{8pt}
\caption*{Table 5. Reference implementation component mapping to PVDM phases.}
\end{table}

An optional business-rules gate (SparkRules, DRL rules run before the Physical phase) can filter or enrich chunks; it is explicitly not a fifth phase. The framework is distributed as the serverless-data-mesh PyPI package and a GHCR container image, with a metadata-driven compiler that expands one mesh YAML into many proof-gated pipelines (bronze, silver, gold) plus durable orchestrators.

The dual clocks are configured through infrastructure parameters: a container clock (lambda\_timeout\_seconds, 1 to 900 seconds, the AWS hard limit), a workload clock (durable\_execution\_timeout\_seconds, default 5400 seconds, up to roughly one year across replays), and a rollback lead time (iceguard\_rollback\_threshold\_ms). The three-account model (Producer, Steward, Publisher) is realized with multi-account Terraform and Lake Formation cross-account grants.

Honest scope note on what the public artifact does. The reference gate (validation/pvdm\_gate.py in \url{https://github.com/vaquarkhan/Proof-gated-publication-PVDM}) is a dependency-free validation model that implements the hardened Verify and Metadata phases. It computes the keyed multiset accumulator of Section 7 (HMAC-SHA256 per element, summed modulo $2^{256}$), and validate\_then\_commit publishes only after it verifies the Steward signature, checks the commit target and an anti-replay nonce, and re-hashes the exact bytes being published against the per-file digests bound at Verify.

It also includes typed canonicalization, N-version and attested transform verification for Profile T, source-to-sink attestation, key-epoch enforcement, and threshold notarization, each mapped to a weakness in the adversarial analysis and exercised by the suite. This gate is deliberately small and stdlib-only so the invariant and its hardening can be tested; it is a reference model, not the production serverless framework. The production framework is a separate repository, is not claimed conformant here, and should not be assumed conformant merely because this reference gate is.

The decision-attestation layer (PVDM-A) and the agentic/MCP integration are described as roadmap and are not shipped; they must not be presented as implemented.

A minimal usage sketch, from the reference implementation, declares the boundary and runs the workload under the coordinator; the outcome is one of committed, rolled back, or verification failed:

\begin{figure}[H]
\centering
\includegraphics[width=0.9\textwidth,keepaspectratio]{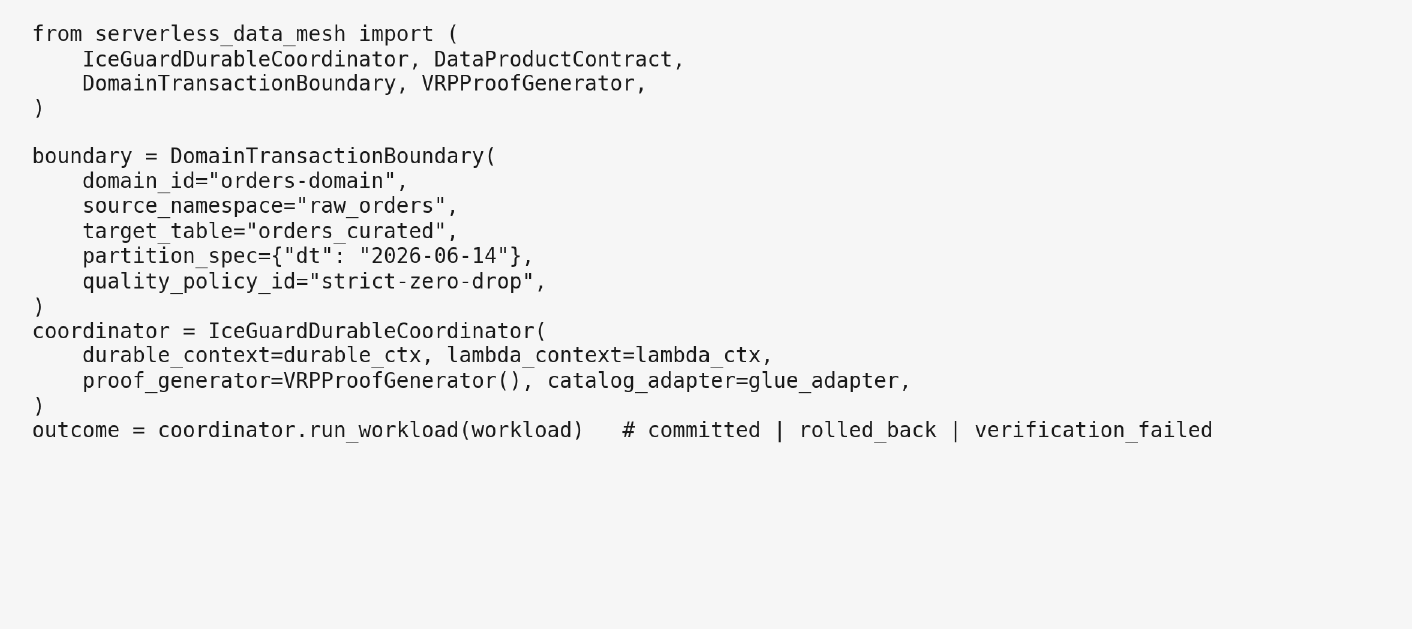}
\vspace{8pt}
\caption*{Figure 6. Listing 3. Minimal usage of the reference implementation.}
\end{figure}

\textit{Runnable reference source: \url{https://github.com/vaquarkhan/Proof-gated-publication-PVDM}.}

\section*{9. Worked Example (Illustrative)}

The following values are illustrative and are used only to make the mechanics concrete; they are not measured results. Consider a backfill of a curated orders partition, split into chunks across several serverless segments.

\begin{itemize}
\item Clean run. The Producer writes each chunk physically; the Steward computes the written multiset digest and compares it to the intended digest declared by the boundary. Digests match, the proof passes, durable replay confirms each chunk exactly once, and the Metadata commit publishes the snapshot. Consumers now see the partition.

\item Silent-drop run. A swallowed exception drops six rows in one chunk. The exit status is success, so a traditional pipeline would commit and analysts would notice drift later. Under PVDM the written multiset digest differs from the intended digest, the proof fails, and by the invariant no consumer-visible snapshot is created; the outcome is verification failed and consumers are unchanged.

\item Retry-duplicate run. A segment times out near the container-clock limit; the watchdog rolls back uncommitted work and the orchestrator resumes. Because completed chunks are replayed rather than rewritten, no duplicate chunk is produced, and the multiset digest would in any case catch a duplicate because it is a multiset, not a set.
\end{itemize}

\section*{10. Evaluation}

\subsection*{10.1 What is already implemented}

The reference implementation ships a consumer-safety benchmark (eval/validate\_then\_commit\_benchmark.py) that injects five fault classes, drop, duplicate, mutation, and schema drift among them, and asserts that every corrupted chunk yields VRP FAIL and therefore never commits a snapshot. A local demonstration (serverless-data-mesh demo) reproduces the gate without cloud infrastructure: a clean write commits and a corrupt write is blocked. The repository also publishes cost estimates comparing Lambda-plus-PVDM to a Glue ETL baseline (for example, roughly 0.025 against 0.44 US dollars at one million rows, and 0.051 against 1.47 at ten million rows).

These are the author's published estimates and a fault-injection safety benchmark; they are cited here as such and are not independent, peer-reviewed production measurements. Beyond this shipped benchmark, Sections 10.3 to 10.5 report reference-gate results measured on cloud hardware (an AWS EC2 c7i.xlarge instance): the adversarial suite, a scaling run of eight thousand fault injections up to one million rows, and a latency breakdown. Those are reproducible microbenchmarks of the gate, still distinct from the production-scale, multi-writer evaluation specified in Section 10.2.

\subsection*{10.2 The protocol for independent, production-scale evaluation}

The protocol below is specified for independent execution at production scale. It generalizes the shipped benchmark and is designed here; production-scale results are future work.

\textbf{Hypotheses.} - H1 (safety): under injected content faults, PVDM produces no consumer-visible snapshot for the faulted chunk, while an exit-code baseline publishes it. - H2 (no false block): under fault-free runs with arbitrary legal file layouts and row orders, PVDM publishes every workload (zero false blocks). - H3 (bounded overhead): PVDM publish latency equals baseline latency plus the multiset-hashing and one extra catalog-commit cost, and this overhead is a small, characterizable fraction of a batch/backfill run. - H4 (concurrent multi-writer safety and liveness): under N producers publishing concurrently to the same table, including writers targeting overlapping partitions, every consumer-visible snapshot still satisfies the invariant, and clean writers are not starved. This is the hypothesis the single-node results of Section 10.8 do not yet test and the one most specific to the federated target environment.

\textbf{Multi-writer concurrency protocol.} Because the target is a federated mesh, the evaluation MUST include a concurrent multi-writer configuration, not only sequential single-writer runs. Concretely: run N independent producers (N in, for example, {2, 4, 8, 16}) that each stage to their own branch and contend on the fast-forward of a shared table; measure fast-forward conflict and retry rates as a function of N and of partition overlap (disjoint partitions versus deliberately overlapping ones); confirm that a failing proof from one writer never blocks or corrupts a concurrent clean writer's publish; and report publish-latency distribution and any starvation under contention. The Iceberg fast-forward is a compare-and-swap on the branch, so a losing writer retries against the new snapshot; the protocol measures whether that retry cost stays bounded and whether the per-chunk invariant composes under concurrency. This is specified here and is the primary subject of the production study.

\textbf{Fault injection.} Silent row drops (remove k rows post-write), partition drops, file truncation, duplicate chunks (replay a chunk without durable dedup), and value corruption (perturb fields). Each fault is applied at a controlled rate.

\textbf{Baselines.} (a) Exit-code publication: commit when the job returns success. (b) Post-load warehouse tests: commit, then run row-count and null-rule checks. (c) WAP with query audits. PVDM is the treatment.

\textbf{Outcome measures.} Detection rate and time-to-detection per fault type; false-positive (false-block) rate on clean runs; publish-latency overhead versus baseline; verification cost as a function of rows and chunk count; and audit re-verification time (offline proof check).

\textbf{Corpus.} Synthetic datasets with known ground truth for controlled fault rates, plus, where available, anonymized production-shaped workloads. Synthetic samples are labeled as demonstrating mechanics, not as evidence of production behavior.

\textbf{Statistics.} Report detection and false-block rates with confidence intervals over repeated trials, and report latency overhead as a distribution (median and tail), not a single number. Pin the multiset-hash construction, the chunking parameters, and seeds so runs reproduce.

\subsection*{10.3 Reference-gate validation results}

Independently of the production framework, I provide a dependency-free reference gate and a thirty-case adversarial and conformance suite [21] that exercises the invariant and every hardening in Sections 12 and 13. I am careful about what this establishes: a passing suite demonstrates that the reference gate behaves as specified against the modeled attacks, which is conformance evidence, not a proof of security or an absence of vulnerabilities, and the suite is authored alongside the design so it reflects the designer's threat enumeration. With that caveat, all thirty checks pass. The cases and outcomes are summarized in Table 3, and grouped by threat family in Figure 4.

To reduce reliance on the designer's own threat enumeration, I complement the enumerated cases with a property-based suite (validation/test\_pvdm\_property.py, using Hypothesis) in which the datasets, the row orders, the file layouts, and the corruptions are machine-generated rather than hand-picked. Four invariants are checked over four hundred generated cases each: soundness (any legal permutation or repartition of the intended multiset verifies PASS, so there are no false blocks), completeness (any generated corruption that changes the multiset, whether a drop, a duplicate, a field mutation, or a composition of these, verifies FAIL and commits nothing), the identity-versus-content split (a value mutation that preserves the identity projection is caught by the content hash), and commit-gate integrity (a replayed nonce is refused). All four properties held across every generated case. Property-based generation explores inputs the author did not enumerate and shrinks any violation to a minimal counterexample, which strengthens the evidence beyond the fixed vectors while still not constituting a proof of security.

\begin{figure}[H]
\centering
\includegraphics[width=0.9\textwidth,keepaspectratio]{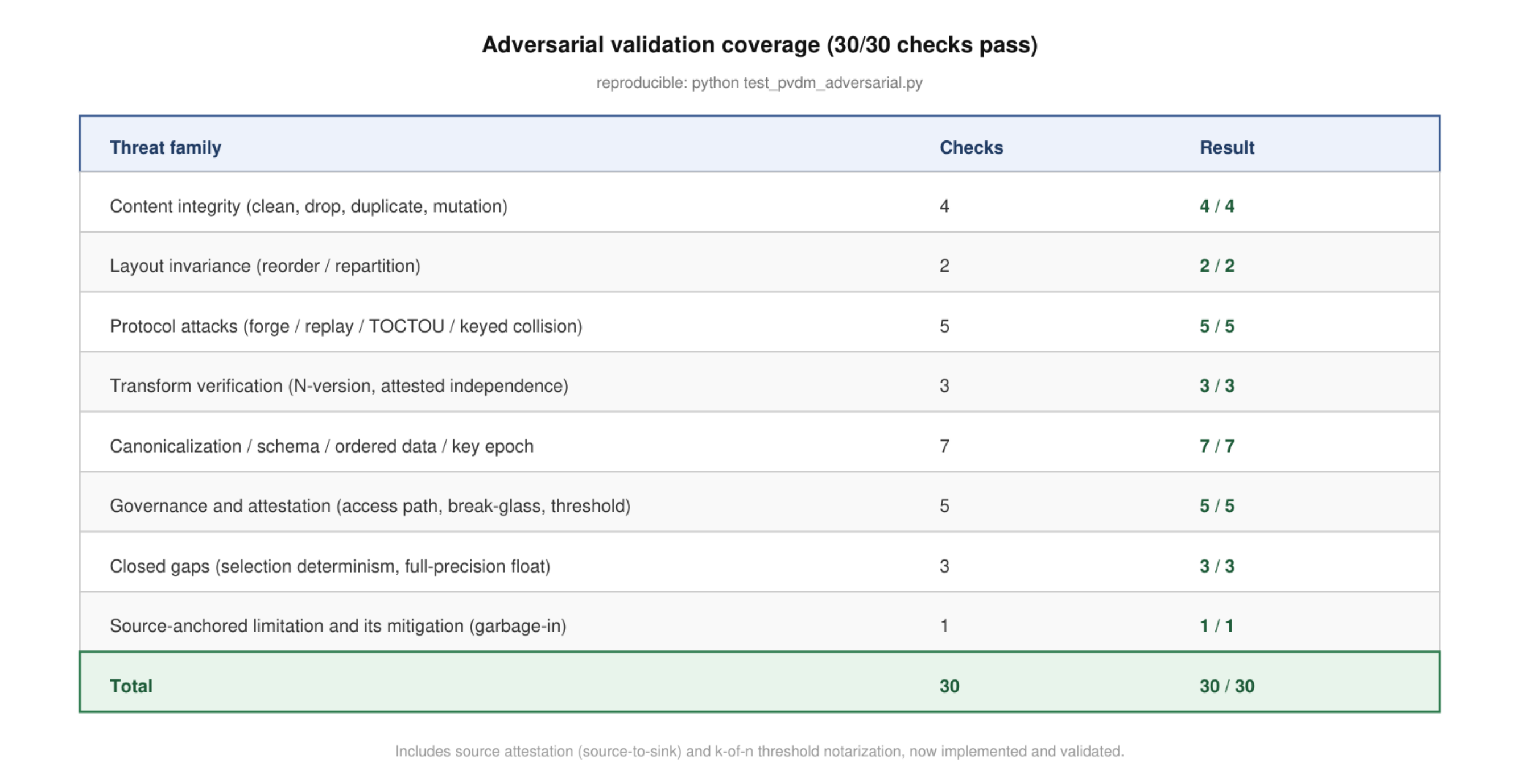}
\vspace{8pt}
\caption*{Figure 7. Adversarial validation coverage grouped by threat family; all thirty checks pass. Reproducible by running the reference suite.}
\end{figure}

\vspace{8pt}\noindent\small\textit{Table 6. Adversarial validation coverage}\vspace{4pt}

\begin{longtable}{|p{0.18\textwidth}|p{0.18\textwidth}|p{0.18\textwidth}|p{0.18\textwidth}|p{0.18\textwidth}|}
\hline
\textbf{\textbf{Case}} & \textbf{\textbf{Weakness}} & \textbf{\textbf{Adversary action}} & \textbf{\textbf{Required outcome}} & \textbf{\textbf{Result}} \\
\hline
clean write & core & none & commit, snapshot visible & pass \\
\hline
reorder rows & L8 & shuffle row order & verify PASS (invariant) & pass \\
\hline
repartition & L8 & different file count & verify PASS (invariant) & pass \\
\hline
drop row & core & remove one row & VRP FAIL, no snapshot & pass \\
\hline
duplicate row & L3 & repeat one row & VRP FAIL (multiset) & pass \\
\hline
mutate value & core & change a field & FAIL via content digest & pass \\
\hline
forged collision & L2 & craft a colliding multiset & keyed digest rejects it & pass \\
\hline
wrong-target replay & L7 & aim proof at another table & commit rejected & pass \\
\hline
nonce reuse & L7 & reuse a valid proof & second commit rejected & pass \\
\hline
post-verify tamper & L4 & change bytes after PASS & commit rejected & pass \\
\hline
forged signature & L7 & flip verdict without key & commit rejected & pass \\
\hline
garbage-in & L1 & corrupt source read & VRP PASS; independent count catches it & pass \\
\hline
transform bug & T & producer miscomputes an aggregate & independent recompute FAILs it & pass \\
\hline
same-impl transform & T & self-compare a transform & certification refused & pass \\
\hline
ordered drop/reshuffle & L11 & drop a middle event; reshuffle & drop FAILs, reshuffle PASSes & pass \\
\hline
canonical coverage & L10 & exclude a consumer field & configuration refused & pass \\
\hline
unsigned override & L13 & silent ``just publish it'' & rejected; signed break-glass logged & pass \\
\hline
access path & L9 & read staged data pre-commit & not visible until committed & pass \\
\hline
key epoch & L14 & change epoch after signing & signature invalidated & pass \\
\hline
typed canonicalization & F2 & raw float in hashed field & rejected; decimals deterministic & pass \\
\hline
functional coverage & F8 & rely on name-only coverage & covered field must change digest & pass \\
\hline
schema fingerprint & F11 & compare across schema versions & fingerprint diverges & pass \\
\hline
partial epoch combine & F9 & mix key epochs in partials & refused to combine & pass \\
\hline
attested independence & F8 & relabel one job as two & same code hash refused & pass \\
\hline
canonical selection & GAP1 & nondeterministic dedup/top-N & shared canonical rule makes impls agree & pass \\
\hline
full-precision float & GAP2 & sub-scale float change (Profile A) & byte-exact hashing catches it & pass \\
\hline
dedup total-order & GAP1b & rows tie on declared order field & full-row tiebreak keeps it deterministic & pass \\
\hline
source attestation & L1 & corrupt source read VRP would accept & independent signed source digest catches it & pass \\
\hline
threshold, under quorum & L6 & one compromised notary tries to publish & rejected below k-of-n & pass \\
\hline
threshold, quorum met & L6 & k distinct valid notaries endorse & publishes only at quorum & pass \\
\hline
\end{longtable}

The suite serves two purposes. First, it turns the safety argument from prose into executable evidence: a reader can run it and confirm that each attack produces the claimed outcome. Second, it encodes the scope boundary directly, since the garbage-in case is written to pass only when the independent source check catches what VRP structurally cannot, so the limitation is represented in the test rather than only in text.

\subsection*{10.4 Reference-gate scaling microbenchmark}

The adversarial suite establishes that each fault class is caught in principle. A separate microbenchmark (validation/benchmark\_aws.py) asks whether detection stays complete as the record count grows, and at what verification cost. For each of four record counts I build a clean baseline table, then in each of two thousand trials inject exactly one single-row fault chosen at random from drop, duplicate, and mutation, and record whether the keyed multiset proof catches it (a caught fault means the digests diverge and the commit fails closed).

I also confirm that a clean sink presented in reversed and randomly permuted physical order still verifies, to count any false block. The run is deterministic under a fixed seed. Because the digest is a sum of per-row element hashes (the incrementality property), each single-row fault is scored in constant time after the rows are hashed once, which is what lets me run two thousand trials per record count; a consistency gate separately confirms that this incremental score equals a full recomputation over the corrupted multiset.

The run reported here executed on an AWS EC2 c7i.xlarge instance (four vCPUs, Amazon Linux 2023, CPython 3.9), and results are in Table 4.

\begin{table}[H]
\centering\small
\begin{tabular}{|p{0.13\textwidth}|p{0.13\textwidth}|p{0.13\textwidth}|p{0.13\textwidth}|p{0.13\textwidth}|p{0.13\textwidth}|p{0.13\textwidth}|}
\hline
\textbf{Records} & \textbf{Fault trials} & \textbf{Detected} & \textbf{Escape rate} & \textbf{Caught by identity} & \textbf{Caught by content only} & \textbf{False-block rate} \\
\hline
1,000 & 2,000 & 2,000 & 0.00\% & 1,310 & 690 & 0.00\% \\
\hline
10,000 & 2,000 & 2,000 & 0.00\% & 1,300 & 700 & 0.00\% \\
\hline
100,000 & 2,000 & 2,000 & 0.00\% & 1,330 & 670 & 0.00\% \\
\hline
1,000,000 & 2,000 & 2,000 & 0.00\% & 1,350 & 650 & 0.00\% \\
\hline
total & 8,000 & 8,000 & 0.00\% & 5,290 & 2,710 & 0.00\% \\
\hline
\end{tabular}
\vspace{8pt}
\caption*{Table 7. Reference-gate scaling microbenchmark: fault detection versus record count.}
\end{table}

\begin{figure}[H]
\centering
\includegraphics[width=0.9\textwidth,keepaspectratio]{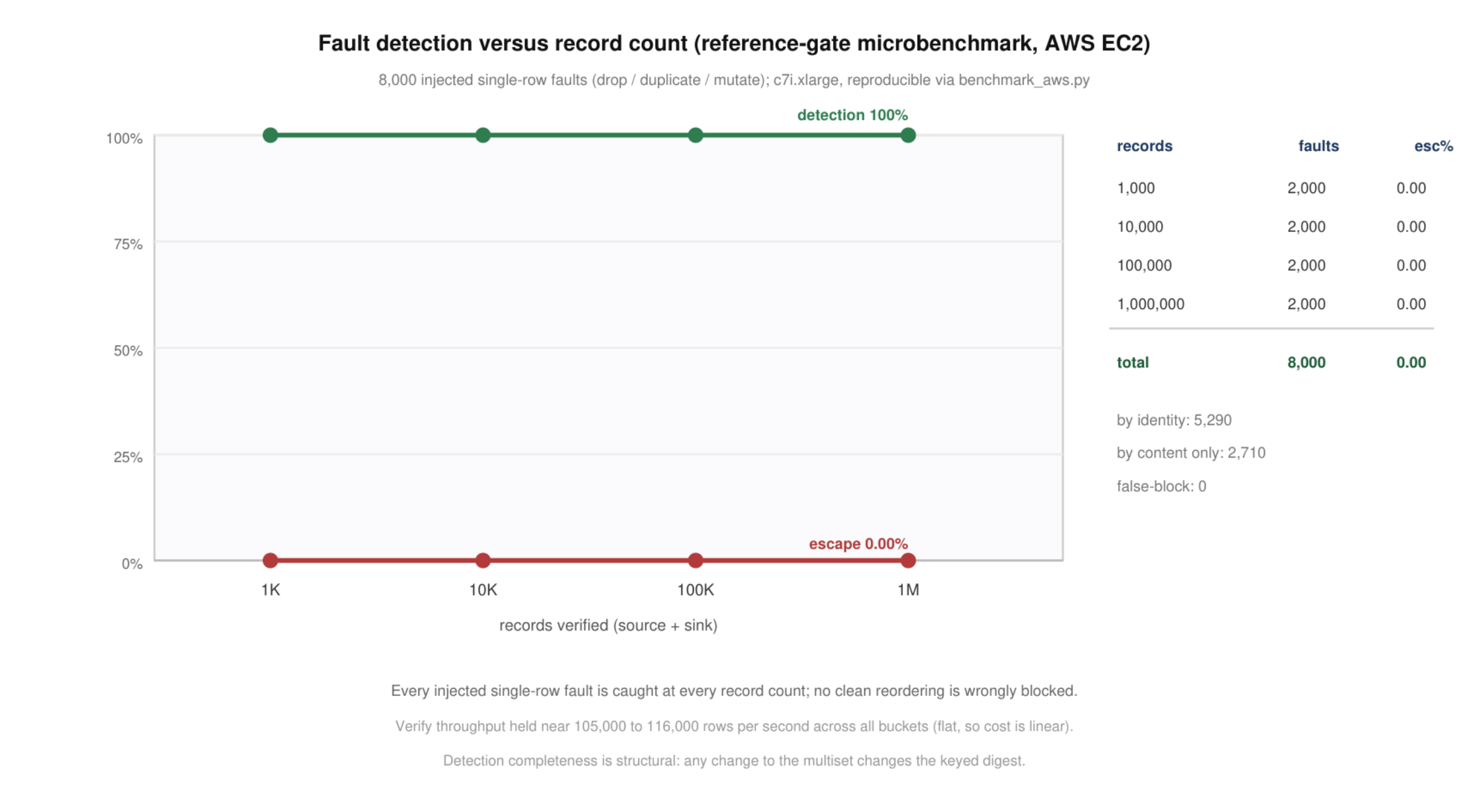}
\vspace{8pt}
\caption*{Figure 8. Detection rate and escape rate versus record count. Detection is 100 percent and escape is 0.00 percent at every record count from one thousand to one million, measured on AWS EC2.}
\end{figure}

The result is not a surprise, and I do not present it as one: detection completeness is structural, because any change to the multiset changes the keyed digest, so the escape rate is zero by construction rather than by luck, and the benchmark confirms that the implementation matches that property at scale. The two rightmost columns also make the diagnostic value concrete: of the eight thousand faults, the identity projection alone flagged the drops and duplicates while value mutations were caught only by the content projection, which is exactly how an operator tells a missing or repeated row apart from a corrupted field without a row-level diff. Verify throughput on this instance held between roughly 105,000 and 116,000 rows per second across the buckets, staying flat as the table grew, which confirms the cost is linear in the record count; I report it as a reference-gate figure on a single instance, not a production performance claim, since a production system would use a vetted hash and parallel hashing. The honest reading of Table 4 is narrow and firm: within the modeled fault classes the gate does not let a corrupted multiset through, and it does not block a clean one, regardless of table size.

\subsection*{10.5 Where the time goes}

To preempt the natural question of what verification costs relative to the rest of a publish, the same benchmark measures the three cost centers on a two hundred thousand row chunk written as sixteen shards, taking the median of five runs on the same c7i.xlarge instance. Staging (serializing the rows to shard files) took about 246 ms, Verify (the identity and content multiset hashing plus per-file digests and the Steward signature) took about 3,796 ms, and the proof-gated Metadata commit (signature check, re-hash of the exact bytes, snapshot write) took about 7 ms, for a total near 4.0 s. In proportion, Verify is about 94 percent of the cost, staging about 6 percent, and the commit gate about 0.2 percent, as shown in Figure 6.

\begin{figure}[H]
\centering
\includegraphics[width=0.9\textwidth,keepaspectratio]{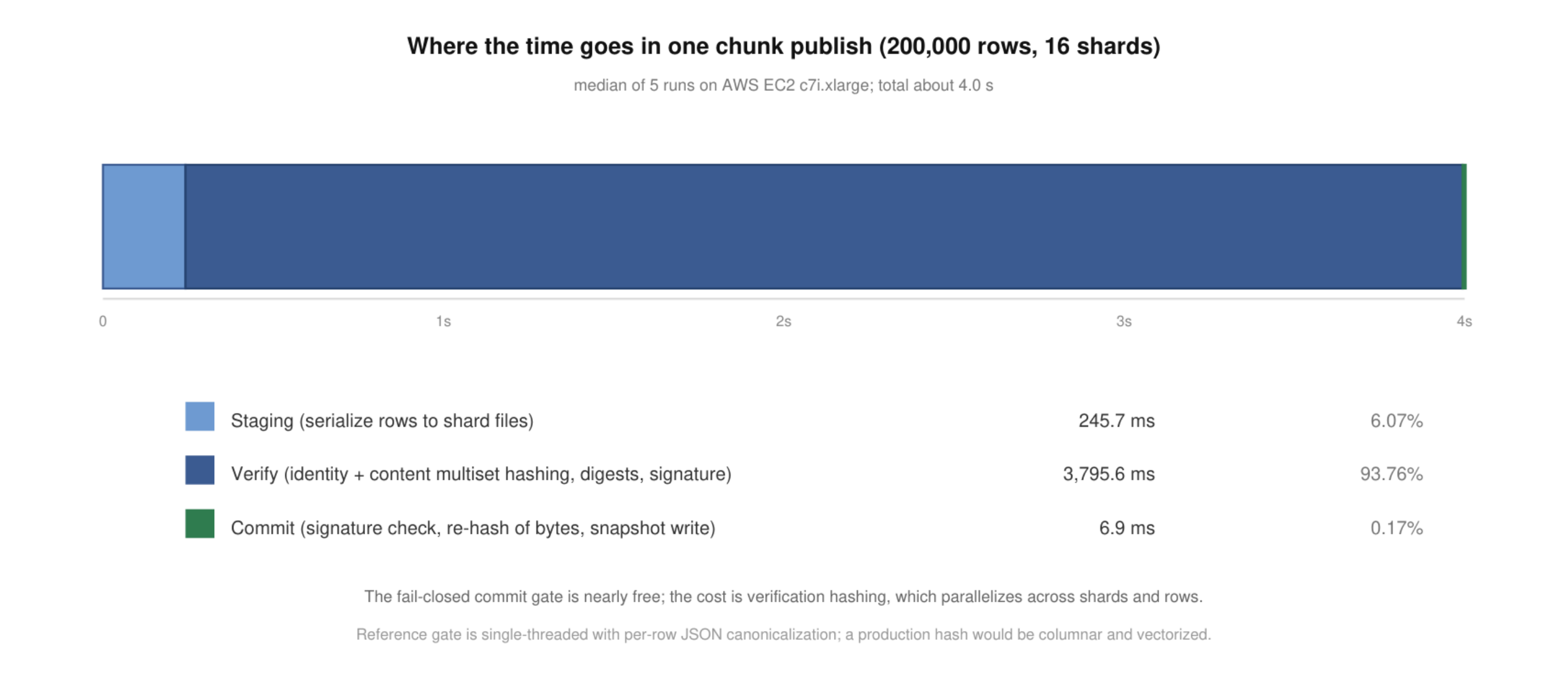}
\vspace{8pt}
\caption*{Figure 9. Latency breakdown of a single chunk publish on a 200,000 row chunk: verification hashing dominates at about 94 percent, staging serialization is about 6 percent, and the proof-gated commit is about 0.}
\end{figure}

Two things follow. First, the fail-closed commit gate itself is nearly free; the guarantee is not paid for at commit time but at verification time. Second, the verification cost is the multiset hashing, which is embarrassingly parallel across shards and rows and, in the stdlib reference gate, runs single-threaded with a per-row JSON canonicalization that a production implementation would replace with a columnar, vectorized hash. The measured split therefore points precisely at the one component worth optimizing, and it does so without changing the correctness result, which does not depend on how fast the hash runs.

\subsection*{10.6 Reproducibility}

Every result in this section is reproducible from the public artifact repository (\url{https://github.com/vaquarkhan/Proof-gated-publication-PVDM}) using only the Python standard library. The adversarial and conformance suite runs as python validation/test\_pvdm\_adversarial.py and prints thirty of thirty checks passing. The scaling and latency results run as python validation/benchmark\_aws.py, deterministic under a fixed seed; the recorded outputs are committed in the repository as validation/aws\_results.json and validation/aws\_results.csv.

The collision study is validation/collision\_analysis.py with validation/collision\_results.json, and the end-to-end Spark and Iceberg outputs are in validation/aws\_spark\_iceberg\_results.json. The runs were executed on an AWS EC2 c7i.xlarge instance (four vCPUs, Amazon Linux 2023, CPython 3.9); the machine descriptor and outputs are committed alongside the code. Reproducing these reference-gate numbers is not a substitute for the independent, production-scale evaluation of Section 10.2.

For a permanent citable record, tag an immutable release and archive it (for example on Zenodo) and cite that DOI from this paper. The reference gate uses only the Python standard library so that an independent reviewer can audit and re-run it in minutes, and the scaling benchmark includes a consistency gate that checks its constant-time fault scoring against a full digest recomputation, so the fast path cannot silently diverge from the specified construction. I make no claim that reproducing these reference-gate numbers substitutes for the independent, production-scale evaluation of Section 10.2; the artifact is offered so that the correctness behavior reported here can be checked by anyone rather than taken on trust.

\subsection*{10.7 Verification-primitive throughput and headroom}

Section 7.5 argues that the throughput of the gate is a property of the chosen hash and canonicalization, not of the protocol. I measured this directly on the same c7i.xlarge instance, computing the content-projection digest over one million rows under four constructions that all preserve the multiset-hash properties (order independence and multiplicity sensitivity were re-checked for each and held). The reference uses a keyed sum of HMAC-SHA256 over per-row JSON canonicalizations; the variants change only the per-row hash, the canonicalization, and the degree of parallelism. Results are in Table 5 and Figure 7.

\begin{table}[H]
\centering\small
\begin{tabular}{|p{0.30\textwidth}|p{0.30\textwidth}|p{0.30\textwidth}|}
\hline
\textbf{Construction} & \textbf{Rows per second} & \textbf{Speedup} \\
\hline
v0 reference (HMAC-SHA256 + JSON canonicalization) & 220,799 & 1.00 \\
\hline
v1 keyed BLAKE2b + JSON canonicalization & 293,831 & 1.33 \\
\hline
v2 keyed BLAKE2b + delimiter canonicalization & 852,230 & 3.86 \\
\hline
v3 v2 parallelized across 4 cores & 1,062,180 & 4.81 \\
\hline
\end{tabular}
\vspace{8pt}
\caption*{Table 8. Verification-primitive throughput and headroom on c7i.xlarge (1M rows).}
\end{table}

\begin{figure}[H]
\centering
\includegraphics[width=0.9\textwidth,keepaspectratio]{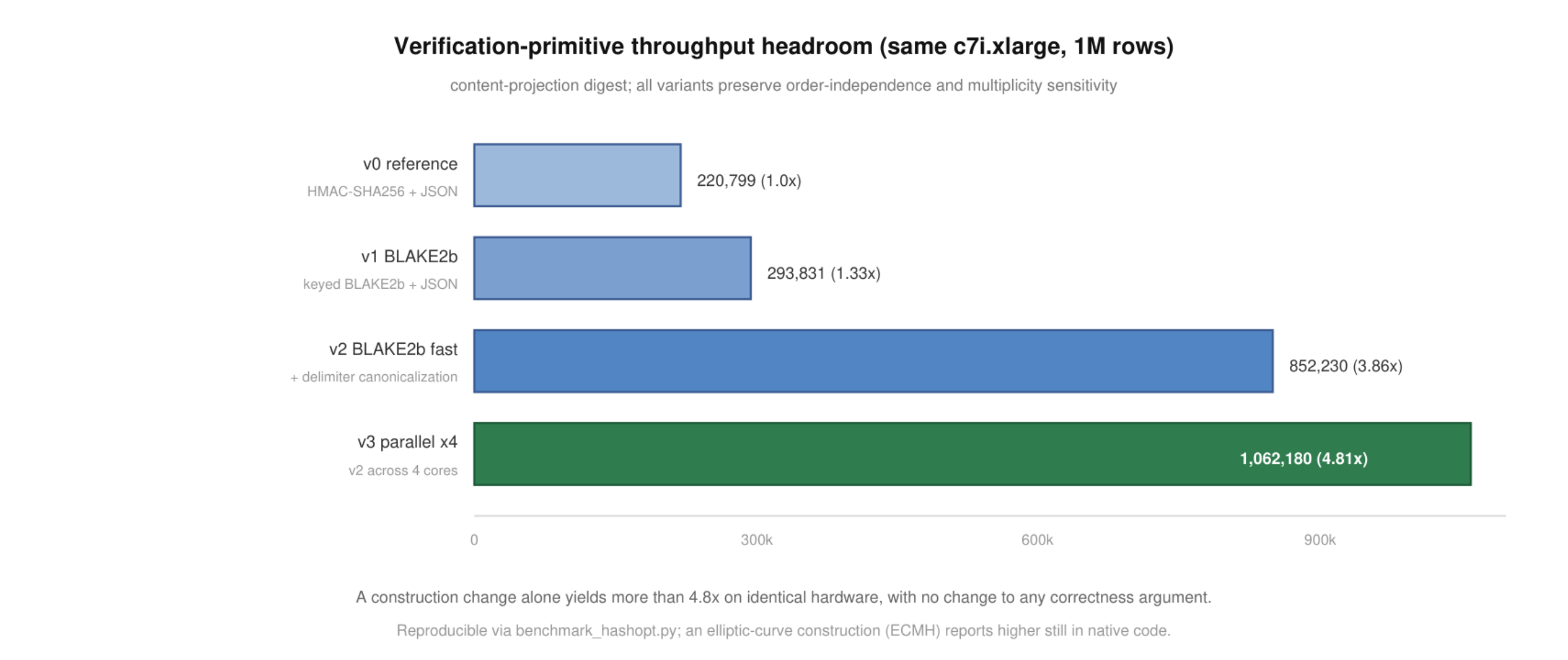}
\vspace{8pt}
\caption*{Figure 10. Content-projection digest throughput for four constructions on the same instance: the reference at about 221,000 rows per second rising to about 1.06 million with a faster keyed hash, cheaper canonica}
\end{figure}

Two honest qualifications. First, the four-core parallelization added only about 1.25x over the single-threaded v2, not 4x, because this naive process-pool form serializes row shards to worker processes; a shared-memory or columnar layout would parallelize closer to linearly, and the digest's additivity places no correctness obstacle in the way. Second, an elliptic-curve construction such as ECMH [22] reports throughput well above any of these variants in optimized native code. The measured point is narrow and sufficient: on identical hardware, a construction change alone yields more than a 4.8x improvement with no change to any correctness argument, so the verification cost reported in Section 10.4 is an implementation property with substantial and demonstrated headroom, not a floor imposed by the protocol.

\subsection*{10.8 End-to-end on Apache Spark and Apache Iceberg}

The results above exercise the gate in isolation. To test PVDM on the real substrate it targets, I ran it end-to-end on Apache Spark 4.0.4 (Scala 2.13, Java 17) with Apache Iceberg 1.11.0 on a single AWS EC2 m7i.4xlarge instance (16 vCPUs, 64 GB), publishing a synthetic TPC-H lineitem table at ten, fifty, and one hundred million rows. The Iceberg write-audit-publish branch flow maps directly onto the PVDM phases: Spark writes the chunk to a staging branch (Physical); a keyed multiset digest is computed over the staged rows and compared to the intended digest as a native, distributed Spark aggregation (Verify); and on a passing proof the table is fast-forwarded from the staging branch to main (Metadata).

I take one warmup and three measured trials per scale under a fixed seed, and separately inject a dropped, a duplicated, and a mutated row at each scale to confirm the gate fails closed on the real table. Results are in Table 6.

\begin{table}[H]
\centering\small
\begin{tabular}{|p{0.10\textwidth}|p{0.10\textwidth}|p{0.10\textwidth}|p{0.10\textwidth}|p{0.10\textwidth}|p{0.10\textwidth}|p{0.10\textwidth}|p{0.10\textwidth}|p{0.10\textwidth}|}
\hline
\textbf{Records} & \textbf{Write (p50)} & \textbf{Stage write (p50)} & \textbf{Verify (p50)} & \textbf{Verify (p95)} & \textbf{Fast-forward (p50)} & \textbf{Verify rows/s} & \textbf{Gate overhead} & \textbf{Faults blocked} \\
\hline
10,000,000 & 1.22 s & 1.23 s & 0.43 s & 0.55 s & 0.019 s & 21.5 M/s & 36.9\% & 3/3 \\
\hline
50,000,000 & 5.24 s & 5.24 s & 1.02 s & 1.04 s & 0.017 s & 49.4 M/s & 19.8\% & 3/3 \\
\hline
100,000,000 & 10.13 s & 9.97 s & 1.88 s & 1.94 s & 0.017 s & 53.2 M/s & 18.7\% & 3/3 \\
\hline
\end{tabular}
\vspace{8pt}
\caption*{Table 9. PVDM on Spark 4.0.4 and Iceberg 1.11.0, single m7i.4xlarge node.}
\end{table}

\begin{figure}[H]
\centering
\includegraphics[width=0.9\textwidth,keepaspectratio]{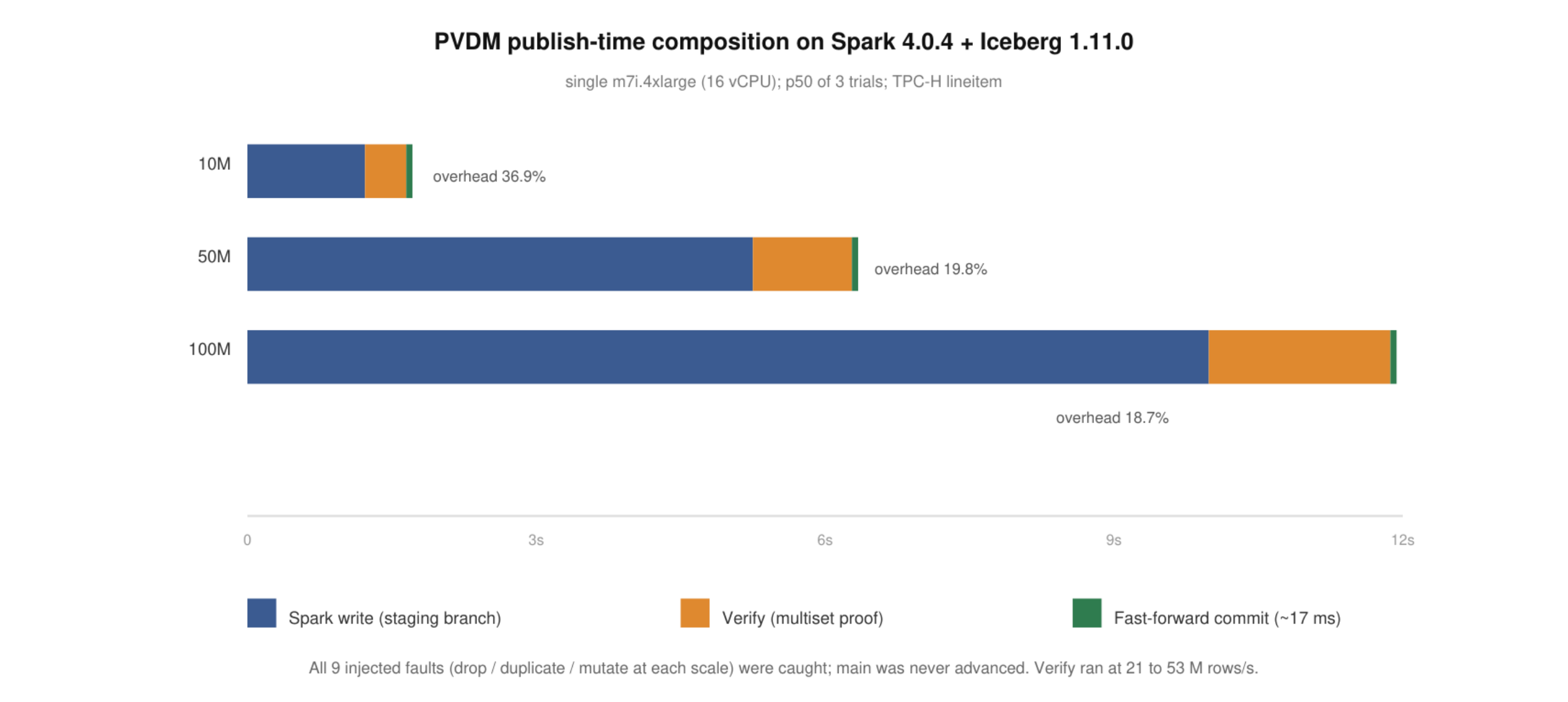}
\vspace{8pt}
\caption*{Figure 11. Publish-time composition on Spark and Iceberg at three scales: the Spark write dominates, the verify step is a modest fraction that shrinks with scale, and the fast-forward commit is negligible.}
\end{figure}

Three results stand out. First, the gate is fail-closed on the real commit path: at every scale, a dropped, duplicated, or mutated row made the staged digest diverge from the intended digest, the fast-forward was withheld, and consumers reading main saw no corrupted snapshot. This is the end-to-end analogue of the reference-gate safety result, now on an actual Iceberg table.

Second, the Metadata commit is nearly free and flat: the fast-forward took about 17 milliseconds regardless of table size, because it is an O(1) metadata operation, confirming on the real format what Section 10.5 measured on the gate. Third, the verify step runs at 21 to 53 million rows per second as a native distributed aggregation, so the total gate overhead is about 19 percent of the write at one hundred million rows and falls as a fraction of the write as scale grows. The verify throughput here is two to three orders of magnitude above the single-threaded standard library gate of Section 10.4, which is the concrete confirmation, on the real engine, of the headroom argued analytically in Section 10.7.

I am deliberately precise about what this measures and what it does not, since overhead is the question an adopter asks first.

\begin{itemize}
\item \textbf{Scope.} The run is single-node and single-writer. It establishes end-to-end correctness and overhead on the real Spark and Iceberg stack, not concurrent multi-writer behavior at production cluster scale, which remains the study specified in Section 10.2.

\item \textbf{Which cost is counted.} The staging write is the same Spark write as the baseline (one to a branch), so the PVDM-specific cost is verify plus fast-forward, reported as the gate overhead. The verify here times a single pass over the staged (written) rows; the intent-side digest is computed during the producer's source read, which the job performs regardless, so it is not counted twice. A standalone verification that hashes both sides from cold would cost about twice the reported verify.

\item \textbf{Hash strength versus the reported number.} The verify in Table 6 uses a fast native additive hash so the digest stays inside the engine, which is why it reaches tens of millions of rows per second. A conformant deployment uses the keyed 256-bit cryptographic multiset hash of requirement N1, whose per-row cost is higher; Section 10.7 characterizes that cost directly (for example a keyed BLAKE2b at roughly 0.85 to 1.06 million rows per second in the reference gate, and SHA-256 is itself a native Spark primitive). The honest consequence is that the roughly 19 percent overhead in Table 6 is a lower bound for the fast hash; the conformant overhead is higher, and I do not claim a measured conformant Spark figure. As a modeled bound rather than a measurement, applying the reference-gate ratio between the conformant and fast constructions (Table 5, roughly a factor of four) to the Table 6 verify fraction would raise the gate overhead at one hundred million rows from about 19 percent to a modeled band of roughly 40 to 75 percent. This is an estimate, not a measured result, and it is an upper bound: Spark computes SHA-256 as a native vectorized primitive whereas the reference-gate ratio is single-threaded Python, so the measured figure is expected to fall below this band. The measured value, and a fully native, vectorized cryptographic multiset-hash operator for Spark, are identified as future work. What Table 6 does establish independent of hash choice is the fail-closed correctness on the real commit path and the O(1) cost of the gated publish, since neither depends on the hash construction.

\item \textbf{Versus the naive alternative.} The same content assurance can be obtained by materializing both sides and computing a full multiset difference, but that requires two shuffle-and-sort passes over the data, whereas the PVDM digest is a single streaming aggregation with an O(1) additive combine. Quantifying that gap at scale is left to the production study; the structural difference (single-pass aggregation versus a distributed anti-join) is why a digest gate is the appropriate mechanism.
\end{itemize}

\subsection*{10.9 Is a twenty to thirty percent overhead acceptable? Cost analysis and mitigations}

I take the strongest form of the objection seriously rather than dismissing it. If PVDM adds twenty to thirty percent to publish time, then a pipeline that runs for four hours pays roughly one extra hour, and for a team on a tight compute budget or a tight window that is not a rounding error. Stated bluntly, the measured overhead is not free and there are workloads for which it is too much. The honest analysis is about what that number actually represents and how far standard engineering brings it down.

First, what the number is. The nineteen to thirty-seven percent of Section 10.8 is the cost of the naive implementation: a second, whole-table pass that reads the staged data back and hashes it, on a single node, after the write. It is an upper bound, not an intrinsic tax, and three properties of the digest let a real deployment reduce it substantially:

\begin{itemize}
\item \textbf{Fuse the proof into the write.} The digest is an additive per-row accumulator, so it can be computed during the write pass instead of in a separate read-back pass. Folding the hash into the job that already scans and writes the rows removes the second I/O pass, which is the dominant part of the measured overhead, leaving only the marginal CPU of hashing. The same applies to the intent side, which can be accumulated during the source read the producer performs regardless.

\item \textbf{Verify only the change, not the table.} Because the digest combines per-chunk partials, an incremental or partitioned write hashes only the new chunk, so the overhead is proportional to the data written, not to table size. For the common case of appending a partition to a large table, the verify cost is a small fraction of a full-table publish.

\item \textbf{Scale the proof out.} The additive combine is embarrassingly parallel, so verification throughput grows with executors while the write is frequently I/O-bound. Section 10.8 already shows the overhead fraction falling from thirty-seven percent at ten million rows to nineteen percent at one hundred million as fixed costs amortize, and a wider cluster continues that trend; Section 10.7 shows a further multiple from the hash construction alone.
\end{itemize}

Second, where the overhead lands. On batch and backfill paths, publish latency is usually off the user-facing critical path, so the wall-clock addition is absorbed by scheduling slack and the real question is compute cost, not user-visible delay. The gated commit itself is nearly free (about seventeen milliseconds, Section 10.8), so the entire cost is the verify pass, which the levers above target directly.

Third, where it is genuinely unacceptable, stated plainly so adopters can self-select out: any sub-second or latency-critical publish path, where even a fused single-pass hash is too much; and pipelines whose compute budget cannot absorb single-digit-percent additional CPU at scale. For those, PVDM is the wrong tool, which is the boundary drawn in Section 11.

The net position, without overclaiming: the twenty to thirty percent figure is the conservative whole-table, two-pass, single-node measurement; a fused, incremental, parallel implementation reduces the effective overhead toward the marginal cost of hashing the newly written data, which for batch and backfill is a cost most integrity-sensitive pipelines can absorb. Measuring the fused and incremental implementations directly is engineering I identify as future work (Section 18); I do not report numbers I have not run.

\section*{11. When to Apply and When Not To}

\textbf{Apply PVDM when} multiple domains publish to a shared Iceberg or Delta lakehouse; when compute is federated across accounts, or is planned to be; when auditors need offline-verifiable, per-chunk evidence that published data matches intent; when backfills run for many minutes across serverless invocations; and when a green job that shipped wrong data has already hurt the organization. The common thread is that the cost of a silently wrong publication is high and the publication cadence is batch or backfill rather than sub-second.

\textbf{Do not apply PVDM when} any of the following holds, because the cost would exceed the benefit:

\begin{itemize}
\item \textbf{Single team, single pipeline, no mesh governance.} With one trusted writer and no federation, the three-account notary adds operational weight without closing a real trust gap; a simpler in-pipeline check suffices.

\item \textbf{Sub-second and per-event streaming hot paths.} PVDM gates at the table-commit boundary and adds a verification cost per commit, so it targets discrete-commit publication: batch, backfill, and micro-batch streaming whose trigger interval is comfortably larger than the per-batch verify time. It is not for low-latency, per-event streaming that must publish in milliseconds, and a streaming exactly-once pipeline the organization already trusts gains little. The dividing line is not batch versus streaming as such but whether publication happens as discrete commits with a latency budget that can absorb a proof; a micro-batch write to an Iceberg or Delta table with second-to-minute triggers can be gated per micro-batch because the verify cost is proportional to the small batch, whereas a sub-second path cannot.

\item \textbf{Producers that will not declare transaction boundaries or intent.} The Verify phase compares against a declared intent; if a domain cannot or will not state what a chunk is, there is nothing to verify against.

\item \textbf{Purely transforming workloads with no independent oracle budget.} For aggregation and join stages the guarantee is only as strong as the independence a deployment funds (Profile T, Section 13.6); a team unwilling to pay for an independent recompute or an authenticated upstream digest gets weak assurance and should not claim the strong guarantee.

\item \textbf{Confidentiality or availability is the actual requirement.} PVDM is an integrity mechanism; it is not encryption, and its fail-closed choice trades availability for safety (Section 12.9). A problem that is really about secrecy or uptime is the wrong fit.
\end{itemize}

\subsection*{11.1 How PVDM changes data-architecture and data-quality practice}

PVDM is not only a mechanism; it proposes a shift in where data quality is enforced. Today, data-quality practice is dominated by post-hoc assertions: tests, expectations, and monitors that run after data is already published and that sample or aggregate rather than prove. They answer ``does the published data look plausible'' after the fact, and they can pass while specific rows are missing. PVDM reframes the question to ``can this data be published at all,'' and answers it with a proof computed before the snapshot becomes visible. This is the same move that transactions made for consistency and that types made for program correctness: turning a property that used to be checked by convention and after the fact into an invariant enforced at the boundary.

Three consequences follow for architecture. First, the definition of data quality gains a new, provable tier: alongside statistical expectations (nulls, ranges, distributions), there is now a binary, cryptographic content-fidelity property (published equals intended) that is either proven or the data does not exist for consumers. Second, accountability in a data mesh becomes structural rather than social: the notary boundary means a domain cannot publish unverified data or quietly delete the evidence of a failure, so governance is enforced by the platform instead of by review. Third, the audit trail becomes an artifact of publication rather than a separate logging concern, because the proof that gated each commit is itself the offline-verifiable record.

I state the case for adoption plainly. PVDM is most likely to become a standard tier of data-quality practice where the failure it prevents is expensive and undetectable by sampling, which is precisely the federated, serverless, open-table-format setting that is now mainstream. It does not replace statistical data-quality testing, which catches semantic problems a content-fidelity proof does not model; the two compose, with content proof as the mandatory gate and quality tests as the advisory layer. The claim is therefore specific: verify-before-commit content integrity belongs in the standard toolkit for lakehouse publication, in the same way write-audit-publish and schema enforcement already do, and PVDM is a concrete, conformance-defined way to provide it.

\section*{12. Adversarial analysis: attack surface and mitigations}

A publication gate is only as strong as its weakest assumption under attack. I enumerate the weaknesses that can be raised against a naive reading of PVDM and specify the mitigation that closes or bounds each. A dependency-free reference gate and an adversarial suite implement and check every row of this table; all checks pass (see validation/).

\vspace{8pt}\noindent\small\textit{Table 10. Attack surface and mitigations}\vspace{4pt}

\begin{longtable}{|p{0.18\textwidth}|p{0.18\textwidth}|p{0.18\textwidth}|p{0.18\textwidth}|p{0.18\textwidth}|}
\hline
\textbf{\#} & \textbf{Weakness (naive PVDM)} & \textbf{Attack} & \textbf{Mitigation} & \textbf{Validated by} \\
\hline
L1 & Intent is the producer's own read & A corrupt source read makes sink == (wrong) source, so VRP passes & Anchor intent to an INDEPENDENT source attestation (source-side row count or signed digest from a different trust domain than the writer); VRP alone proves read-to-sink fidelity, not upstream truth & garbage-in test: VRP passes, independent count catches the drop \\
\hline
L2 & Weak/unkeyed multiset hash & Craft a colliding multiset (drop two rows, add one whose contribution sums to the same digest) & KEYED cryptographic multiset hash (MSet-Add-Hash with HMAC-SHA256 per element, summed mod $2^{256}$); the Steward-held key makes element forgery infeasible & keyed hash defeats a collision an unkeyed sum-hash accepts \\
\hline
L3 & Set semantics hide duplicates & Duplicate a row on an at-least-once retry & Additive multiset semantics: a duplicate adds its element hash twice and changes the digest & duplicate-row test fails closed \\
\hline
L4 & Time-of-check to time-of-use & Overwrite staged files after VRP PASS but before commit & Proof binds per-file content digests; the commit re-hashes the exact bytes being published and refuses on mismatch & post-verify tamper test rejected \\
\hline
L5 & Per-chunk vs per-workload atomicity & Consumers see chunk 1 while chunk 2 is mid-flight & Scope the guarantee per chunk explicitly; for all-or-nothing, stage to an Iceberg branch and fast-forward only when every chunk passes & design requirement (documented) \\
\hline
L7 & Replay / forged proof & Reuse an old PASS proof, or flip verdict without the key, or aim a valid proof at another table & Steward-sign the proof over (workload\_id, chunk\_id, target, partition, digests, nonce); commit checks signature, target, and an anti-replay nonce ledger & forged-proof, wrong-target, and reused-nonce tests all rejected \\
\hline
L8 & Encoding/order sensitivity & Same data in different layout falsely fails, or corruption hides in encoding & Canonical row serialization before hashing gives order and partition invariance & reorder and repartition tests pass \\
\hline
\end{longtable}

The remainder of this section addresses each attack in turn. For every weakness I give the adversary's goal and capability, show why a naive reading of PVDM is vulnerable, state the specific mitigation, argue why the mitigation holds, and name the residual risk that remains after it. Each defense ends with the concrete check in the validation suite that exercises it.

\subsection*{12.1 Intent anchored to the producer's own read (L1)}

\textbf{Adversary and goal.} A faulty or malicious source reader inside the Producer account wants a corrupted partition to be published while VRP still reports PASS. \textbf{Capability.} The Producer both reads the source and writes the sink, and it supplies the ``intended'' row set from which the intent digest is computed.

\textbf{Why naive PVDM is vulnerable.} VRP compares an intent multiset to a written multiset. If the intent multiset is simply the rows the Producer read, then a reader that silently drops rows produces an intent that already omits them; the sink faithfully reflects that same truncated read, so intent equals sink and VRP passes. The proof is sound but establishes a narrower property than it may appear to: it certifies read-to-sink fidelity, not fidelity to the true upstream state. This is the most consequential limitation of the protocol, and it is stated explicitly rather than obscured by the cryptography.

\textbf{Hardening.} Move the anchor of intent outside the writer's trust domain. Three mechanisms compose: (a) require the source system to emit a signed provenance token, minimally an authenticated row count and, where feasible, a source-side multiset digest, computed by the source rather than the reader; (b) run an independent Steward-side reader that recomputes the intent digest from the source under a different identity, so a single compromised Producer cannot dictate both sides of the comparison; and (c) where the source is itself a governed table, chain PVDM proofs so that the intent of a downstream stage is the verified content digest of the upstream stage, making corruption detectable at the first governed hop.

\textbf{Why it holds and residual risk.} With an independent intent anchor, a Producer-side read fault now produces intent (from the source or Steward reader) that disagrees with the truncated sink, and VRP fails closed. The residual risk is genuine and bounded: if the true source is itself corrupt, or the source cannot attest, PVDM certifies fidelity to the best available statement of intent and no further. I therefore state the guarantee precisely as read-to-sink fidelity by default, upgraded to source-to-sink fidelity exactly when an independent source attestation is available. \textbf{Validation.} Two tests establish both halves: the garbage-in test shows VRP passing on a corrupt read while an independent count catches it (the limit), and the source-attestation test implements a signed source attestation (an authenticated row count and content digest from a domain distinct from the writer) and shows that the same corrupt read is now caught, with a forged attestation rejected. The source-to-sink upgrade is therefore implemented and validated in the reference gate, not merely proposed.

\subsection*{12.2 The multiset hash must be keyed (L2)}

\textbf{Adversary and goal.} A Producer that knows the hashing scheme wants to construct a corrupted multiset whose digest collides with the intended digest, so that a dropped-and-substituted set of rows verifies. \textbf{Capability.} Full knowledge of the hash function and its parameters; ability to choose the rows written.

\textbf{Why naive PVDM is vulnerable.} A tempting implementation uses an unkeyed homomorphic or additive digest, for example the modular sum of an unkeyed per-row hash, or worse a sum over a numeric column. Such a digest is linear and public, so an adversary can solve for a substitution: drop two rows and add one whose contribution equals their sum, and the digest is unchanged. An XOR-based set hash is even weaker, ignoring multiplicity entirely. The elegance of an order-independent digest becomes a liability if the per-element contribution is predictable.

\textbf{Hardening.} Use a keyed cryptographic multiset hash. The reference gate computes, for each canonicalized row, an element hash HMAC-SHA256 under a Steward-held key, then sums the element hashes modulo a large modulus. The additive combination preserves order independence and multiplicity sensitivity; the keyed element hash makes the per-row contribution a pseudorandom value the adversary cannot predict or invert without the key.

\textbf{Why it holds and residual risk.} Under the standard assumption that HMAC-SHA256 is a pseudorandom function, finding two distinct multisets with equal keyed sum is infeasible for an adversary without the key; this is the incremental-multiset-hash security argument applied to the additive construction. The residual risk is key management: the security reduces to protecting the Steward key. It must live in a key-management service, never in the Producer account, and rotate on a schedule; a leaked key reintroduces the forgery. \textbf{Validation.} The collision test constructs a substitution that an unkeyed sum accepts and shows the keyed multiset hash rejects it.

\subsection*{12.3 Duplicates must not be absorbed (L3)}

\textbf{Adversary and goal.} An at-least-once delivery path, or a retried chunk, duplicates rows; the Producer wants the duplicate to pass unnoticed. \textbf{Capability.} Ordinary retry behavior, no special power required; this is as much a correctness hazard as an attack.

\textbf{Why naive PVDM is vulnerable.} If the digest is a set hash, duplicated rows collapse to the same set and the duplication is invisible. Set semantics are a common default and are wrong here, because a duplicated financial row is a real corruption.

\textbf{Hardening.} Use multiset (additive) semantics: a duplicated row contributes its element hash a second time, changing the digest. The intent multiset carries the intended multiplicity, so any change in multiplicity, up or down, diverges.

\textbf{Why it holds and residual risk.} Additive accumulation is exactly multiplicity-preserving, so a duplicate is detected identically to any other divergence. The residual concern is legitimate duplicates in append-only sources; these must be modeled in the intent (the intent multiset includes them) rather than suppressed, and the pattern documents that dedup policy is a property of the declared intent, not of the gate. \textbf{Validation.} The duplicate-row test fails closed.

\subsection*{12.4 Time-of-check to time-of-use (L4)}

\textbf{Adversary and goal.} A Producer that owns the Physical staging area wants to pass VRP on clean bytes and then swap in corrupt bytes before the catalog commit publishes them. \textbf{Capability.} Write access to the staged files between the Verify and Metadata phases.

\textbf{Why naive PVDM is vulnerable.} If Verify hashes a staging path and Metadata later publishes whatever is at that path, there is a window in which the bytes can change. The proof would then certify content that is no longer the content being published, defeating the gate.

\textbf{Hardening.} Bind the proof to content, not to a path. The proof records a per-file content digest of the exact bytes verified. The Metadata gate re-hashes the exact objects it is about to publish and refuses to commit unless every digest matches the proof. In production this is reinforced by publishing immutable object versions and referencing those version identifiers in the commit, so that the verified bytes and the published bytes are provably identical objects.

\textbf{Why it holds and residual risk.} Any post-verification mutation changes at least one file digest, so the re-hash at commit diverges and the commit is rejected fail-closed. The residual cost is a second hash pass at commit time over the file digests, which is negligible relative to the write. \textbf{Validation.} The post-verify tamper test mutates a file after PASS and shows the commit is rejected.

\subsection*{12.5 Per-chunk versus per-workload atomicity (L5)}

\textbf{Adversary and goal.} Not an attacker so much as a correctness gap: a consumer must not see a partially published workload (chunk one visible while chunk two is still in flight). \textbf{Capability.} Normal multi-chunk execution.

\textbf{Why naive PVDM is vulnerable.} The invariant is stated per chunk. If each chunk commits independently, a workload that must be all-or-nothing leaks its early chunks before the late ones are verified, which for some consumers is itself a correctness violation.

\textbf{Hardening.} Make the atomicity scope explicit and configurable. For per-chunk workloads the default stands. For workload-level atomicity, stage all chunks to an isolated table branch, verify every chunk, and fast-forward the branch into the consumer-visible table in a single catalog operation only after all proofs pass; a single failing chunk discards the branch. This reuses the table format's atomic branch publish as the commit primitive.

\textbf{Why it holds and residual risk.} Branch fast-forward is atomic at the catalog, so consumers transition from the old snapshot to the fully verified new one with no partial state. The residual cost is retaining the branch and its staged data until the last chunk verifies, which increases peak storage and delays visibility for long workloads. \textbf{Validation.} This is a design requirement rather than a unit-tested property in the reference gate; the gate validates the per-chunk invariant that the branch strategy composes.

\subsection*{12.6 Insider and notary compromise (L6)}

\textbf{Adversary and goal.} A compromised Steward, or an insider with Steward privileges, wants to bless corrupt content or forge a passing proof. \textbf{Capability.} Control of the notary that signs proofs and performs the commit.

\textbf{Why naive PVDM is vulnerable.} The Steward is a trust anchor; if it is fully compromised, it can sign a PASS for anything. No single-notary scheme escapes this entirely.

\textbf{Hardening.} Reduce and split trust. Separate the Producer, Steward, and Publisher accounts so that no single principal both writes data and attests to it; store proofs and checkpoints in the Steward where the Producer cannot delete them; and, for high-assurance deployments, require multiple independent attestations (for example a second verifier in a different account or a threshold signature) so that a single compromised notary is insufficient to publish. Proofs are append-only and externally auditable, so a compromise is at least detectable after the fact.

\textbf{Why it holds and residual risk.} Splitting duties converts a single point of failure into a collusion requirement, and offline-verifiable proofs make tampering evident to an auditor. The decentralization axis is explicit and can be pushed further: a threshold or multi-party notary distributes the signing authority so that no single account can publish, and publishing the append-only proof stream to an externally auditable transparency log, in the manner of Certificate Transparency [19], lets any party detect a notary that signs inconsistent proofs. These reduce, but do not remove, the trust placed in the notary set. The residual risk is irreducible: a coordinated compromise of the required set of trust anchors defeats any attestation scheme, so the goal is to raise the bar and preserve detectability, not to claim impossibility. \textbf{Validation.} The forged-proof test shows that without the Steward signing key an attacker cannot produce a proof the commit gate accepts, and two threshold-notarization tests implement k-of-n attestation and show that a single notary cannot meet a two-of-three threshold while two distinct valid notaries can, so the multi-party mitigation is implemented and validated in the reference gate rather than only described.

\subsection*{12.7 Replay, forgery, and misdirection (L7)}

\textbf{Adversary and goal.} A Producer wants to reuse a stale PASS proof for a new corrupt commit, aim a valid proof at a different table, or forge a PASS without the notary. \textbf{Capability.} Access to past proofs and to the commit API.

\textbf{Why naive PVDM is vulnerable.} If a proof is just a verdict, an old PASS can be replayed, a proof for table A can be presented for table B, and a verdict flipped from FAIL to PASS if the proof is unsigned.

\textbf{Hardening.} Bind and sign. Each proof is Steward-signed over its full body, which includes the workload identifier, chunk identifier, target table, partition, the per-file digests, and a fresh nonce. The commit gate verifies the signature, checks that the target matches the intended commit, and consults an anti-replay nonce ledger so a nonce is usable once. A flipped verdict invalidates the signature; a wrong target is detected by the binding; a replayed nonce is rejected by the ledger.

\textbf{Why it holds and residual risk.} Signature verification reduces forgery to key compromise (see L6); the target binding prevents misdirection; the nonce ledger prevents replay. The residual concern is ledger durability and scope: the nonce set must be persistent and consistent across the Steward's commit path, or replay protection weakens; this is a standard exactly-once ledger engineering requirement. \textbf{Validation.} Three checks cover this: a forged proof with a bad signature is rejected, a valid proof aimed at the wrong target is rejected, and a reused nonce is rejected on the second commit.

\subsection*{12.8 Canonicalization and non-determinism (L8)}

\textbf{Adversary and goal.} Two failure modes: corruption that hides in encoding differences, and false failures when identical data is serialized differently by two engines. \textbf{Capability.} Normal heterogeneity of readers, writers, and engines.

\textbf{Why naive PVDM is vulnerable.} If rows are hashed in their raw serialized form, then column reordering, differing null encodings, decimal scale, timezone normalization, or floating-point representation cause the intent and sink digests to differ for logically identical data, producing false FAILs; conversely, sloppy canonicalization could let a real change hide in a field that is normalized away.

\textbf{Hardening.} Define a canonical serialization applied identically on both sides before hashing: sorted field names, normalized types, an explicit null representation, fixed decimal scale, and UTC timestamps. Non -deterministic derived fields (generation timestamps, engine-specific identifiers) are either excluded from the content projection or canonicalized to their deterministic form, and the pattern verifies the deterministic projection of the data.

\textbf{Why it holds and residual risk.} Identical logical data canonicalizes to identical bytes, giving order and layout invariance and eliminating false FAILs, while any change to a verified field still diverges. The residual risk is scope: fields excluded as non-deterministic are outside the guarantee, so the canonical projection must be chosen so that everything a consumer relies on is inside it. \textbf{Validation.} The reorder and repartition tests confirm that layout changes preserve PASS, and the mutation test confirms a changed value still fails.

\subsection*{12.9 Availability and the fail-closed tradeoff (L12)}

\textbf{Adversary and goal.} An adversary who cannot forge a PASS may instead try to induce FAILs to deny publication, turning a safety mechanism into an availability weapon. \textbf{Capability.} Ability to perturb the write path or inject transient mismatches.

\textbf{Why this matters.} Fail-closed means a mismatch blocks publication. That is the desired safety behavior, but it makes denial of publication the natural attack, and it raises the operational question of what happens to staged data and to the workload on a FAIL.

\textbf{Hardening.} Pair fail-closed safety with bounded, audited recovery: on FAIL the Physical phase rolls back uncommitted files (no orphans), the workload records a verification-failed outcome with the proof retained as evidence, and an operator-visible repair path re-reads and re-verifies rather than blindly retrying. Rate and anomaly monitoring on FAILs surfaces a denial attempt.

\textbf{Why it holds and residual risk.} The system prefers a blocked, observable, recoverable state to a silent corruption, which is the correct tradeoff for regulated data. The residual risk is that a determined denial-of-publication adversary can delay data; this is an availability cost consciously accepted in exchange for never publishing wrong data, and it is why the pattern is scoped to batch and backfill rather than latency-critical paths.

On the expected FAIL rate under normal operation, the design intent is that a conformant clean write never fails: by the liveness property, if the written multiset equals the intended multiset the proof passes, so the false-block rate under fault-free runs is zero (validated at zero across 8,000 trials in Section 10.4 and across the multi-scale runs in Section 10.8). A FAIL under normal operation therefore signals a genuine content fault, not routine noise, which is what makes a FAIL an actionable alert rather than a nuisance. The one source of spurious FAILs is a canonicalization or profile misconfiguration (for example an undeclared locale or tie-break in a Profile T stage), which is a configuration error surfaced at onboarding rather than a steady-state rate; the conformance profiles and the typed canonicalization of Appendix A exist precisely to drive that source to zero before a stage is certified.

\subsection*{12.10 The transformation-oracle problem (T)}

\textbf{The principal limitation.} The multiset comparison D(intent) = D(written) is only meaningful when an intent exists independently of the producer. For a row-preserving stage, ingest, format conversion, or a filter with a known predicate, the intent is the source, so VRP is meaningful. For a transforming stage, an aggregation, a join, or a dedup, the output multiset is not the input multiset, so there is no pre-existing intent to compare against; the only way to know the correct output is to compute the transform, which is exactly what the producer did. Comparing the producer's output to itself proves write-fidelity, not compute-correctness, and this is a genuine limit that no amount of hashing removes.

\textbf{Hardening (make it normative, not hidden).} PVDM defines a conformance profile per stage. Profile A (row-preserving) verifies output against the source. Profile T (transforming) requires an independent oracle: either an authenticated upstream digest chained from a prior verified stage, or N-version recompute in which a second, independently implemented job recomputes the output and its digest must equal the producer's. The gate refuses to certify a Profile T stage whose two sides used the same implementation, because that provides no assurance about compute correctness.

\textbf{Why it holds and residual risk.} With an independent implementation, a producer compute bug yields an output that disagrees with the oracle and the gate fails closed; the validation suite shows an independent recompute catching an undercount and shows the same-implementation case being refused rather than falsely passed. The residual cost is substantial: Profile T doubles compute for the verified stages that opt into N-version recompute, or requires an attestable upstream. This is the cost of verifying transformations, and the specification states it explicitly rather than implying content correctness without additional work.

\subsection*{12.11 Ordered data (L11)}

\textbf{Attack.} For data whose order is semantically meaningful (event sequences, slowly changing dimensions), a multiset digest is order-independent by design and cannot detect a reordering that matters. \textbf{Hardening.} Profile O requires an explicit sequence or version field folded into the identity projection; the digest then detects any add, drop, duplicate, or mutation, and logical order is recovered by sorting on that field while physical reshuffling stays correctly invariant. \textbf{Residual.} Order must be represented as data; PVDM does not verify implicit physical order, and the standard requires ordered datasets to carry the field. \textbf{Validation.} The ordered-data test shows a dropped middle event caught and a physical reshuffle tolerated.

\subsection*{12.12 Verification through the same extractor and canonicalizer (L10)}

\textbf{Attack.} VRP hashes rows it extracts from the files; a buggy extractor that silently skips a row group, or a canonicalization that normalizes away a consumer-relevant field, could let both sides agree on wrong content. \textbf{Hardening.} Verification MUST read rows the same way consumers will, so the extraction under test is the extraction in production; and a canonical-coverage guard refuses any configuration that excludes a consumer-relevant field from the hashed content projection. Canonicalization is treated as a security-critical, version-pinned, tested component. \textbf{Residual.} Fields deliberately outside the content projection are outside the guarantee, so the projection must cover everything consumers rely on. \textbf{Validation.} The coverage-guard test refuses excluding a relevant field.

\subsection*{12.13 No silent override (L13)}

\textbf{Attack.} Under deadline pressure a team adds a ``just publish it'' bypass, silently voiding the guarantee; this operational escape hatch is the real-world weakest link. \textbf{Hardening.} There is no silent override. The only sanctioned bypass is a break-glass commit that requires a signature from a break-glass key distinct from the Steward key, records an audit event naming the operator and reason, and consumes a nonce; an unsigned or misdirected override is rejected. \textbf{Residual.} A break-glass event is a deliberate, logged, attributable risk acceptance, not a hole; monitoring its rate is a control. \textbf{Validation.} The override test rejects an unsigned bypass and logs a signed one.

\subsection*{12.14 Access-path integrity and key epochs (L9, L14)}

\textbf{Attack (L9).} The invariant ``no consumer-visible snapshot on FAIL'' holds only if every consumer read resolves to a gated, committed snapshot. Three concrete paths could bypass that: a time-travel or version-as-of read that names a snapshot which existed on the staging branch but was never fast-forwarded to main; a direct read of the staging branch or of the underlying object paths, sidestepping the catalog entirely; and a stale metadata cache or pinned metadata pointer that continues to serve a pre-gate state. \textbf{Hardening.} Each path is closed by an explicit normative requirement, and the requirement names where the control lives because it is partly outside the protocol. Consumer identities MUST be granted read access only to main (or to explicitly published branches), never to staging branches or to the raw object prefixes, which is enforced by catalog and object-store access control rather than by the gate.

Time-travel MUST be restricted to snapshots reachable from a gated commit; a staging-branch snapshot that never fast-forwarded is not consumer-addressable, which the branch model already provides because a failed proof performs no fast-forward and leaves main unchanged. Metadata caches and pinned pointers MUST honor snapshot expiry and MUST NOT be refreshed from a branch a consumer cannot read. I state plainly that this is a shared responsibility: the gate guarantees that a failed proof produces no committed main snapshot, but preventing a consumer from reaching around the catalog to staged bytes is an access-control obligation on the deployment, and a deployment that grants consumers direct branch or object access voids the invariant regardless of the gate. \textbf{Attack (L14).} Key rotation mid-workload could let partial digests computed under different keys combine incorrectly, or an attacker could claim a different epoch. \textbf{Hardening.} The proof binds a key epoch inside the signed body, partial digests MUST share an epoch to combine, and changing the epoch invalidates the signature. \textbf{Validation.} The access-path test shows staged data invisible until commit; the key-epoch test shows the epoch is inside the signed body.

\section*{13. Scope, conformance profiles, and normative requirements}

For PVDM to serve as a standard rather than a technique, its guarantee must be stated precisely and its requirements must be normative. This section does that. It uses the keywords MUST, MUST NOT, and SHOULD in the sense of RFC 2119 [8].

\subsection*{13.1 The exact guarantee}

PVDM guarantees, for a chunk published under it, that the consumer-visible content equals a declared intent whose fidelity depends on the stage profile, that the published bytes are exactly those verified, and that the certifying proof cannot be forged, replayed, or misdirected without compromising the Steward key. It does not guarantee upstream truth beyond the declared intent, confidentiality, or availability; the last is consciously traded for fail-closed safety. This scoped statement is the claim; anything broader is not claimed.

\subsection*{13.2 Conformance profiles}

\begin{itemize}
\item \textbf{Profile A (row-preserving).} The stage does not change the row multiset (ingest, format conversion, known-predicate filter). Intent is the source read; an independent source attestation SHOULD anchor it, and MUST anchor it for regulated data, to upgrade read-to-sink fidelity to source-to-sink fidelity.

\item \textbf{Profile T (transforming).} The stage changes the multiset (aggregation, join, dedup). The output MUST be verified against an independent oracle: an authenticated upstream digest chained from a prior verified stage, or an N-version recompute by an independently implemented job. A Profile T certification whose two sides share an implementation MUST be refused.

\item \textbf{Profile O (ordered).} The dataset has semantically meaningful order [16]. It MUST carry an explicit sequence or version field, which MUST be included in the identity projection.
\end{itemize}

A pipeline declares a profile per stage; a stage that does not meet its profile's requirement MUST NOT be certified as conformant.

\subsection*{13.3 Normative requirements}

\vspace{8pt}\noindent\small\textit{Table 11. Conformance profiles}\vspace{4pt}

\begin{longtable}{|p{0.30\textwidth}|p{0.30\textwidth}|p{0.30\textwidth}|}
\hline
\textbf{\#} & \textbf{Requirement} & \textbf{Level} \\
\hline
N1 & The content digest MUST be a keyed cryptographic multiset hash; unkeyed or homomorphic sums MUST NOT be used & MUST \\
\hline
N2 & The Steward key MUST reside in a key-management service and MUST NOT be readable by the Producer & MUST \\
\hline
N3 & The digest MUST use multiset (multiplicity-preserving) semantics, not set semantics & MUST \\
\hline
N4 & The proof MUST bind per-file content digests, and the commit MUST re-verify the exact published bytes & MUST \\
\hline
N5 & The proof MUST be Steward-signed and MUST bind workload, chunk, target, partition, key epoch, and a nonce; the commit MUST enforce target match and a durable, linearizable anti-replay nonce ledger [18] & MUST \\
\hline
N6 & Rows MUST be canonicalized identically on both sides before hashing; the content projection MUST cover every consumer-relevant field & MUST \\
\hline
N7 & Verification MUST read rows using the same extraction path consumers use & MUST \\
\hline
N8 & Transforming stages MUST verify against an independent oracle; same-implementation self-comparison MUST NOT be certified & MUST \\
\hline
N9 & Ordered datasets MUST carry an explicit sequence field in the identity projection & MUST \\
\hline
N10 & On FAIL the system MUST NOT publish and MUST roll back staged files; any override MUST be a signed, logged, attributable break-glass event & MUST \\
\hline
N11 & Consumer reads MUST be confined to the gated catalog snapshot; staged branches MUST NOT be consumer-readable & MUST \\
\hline
N12 & The Producer, Steward, and Publisher SHOULD be separate trust domains; high-assurance deployments SHOULD require multiple independent attestations & SHOULD \\
\hline
N13 & Verification MUST cover the full write; sampling MUST NOT be used when the guarantee is claimed & MUST \\
\hline
N14 & Rows MUST be canonicalized per the typed rules of Appendix A; raw binary floats MUST NOT appear in a hashed projection (carry them as fixed-scale decimals) & MUST \\
\hline
N15 & The proof MUST bind a schema fingerprint and an immutable source snapshot identifier; comparisons across differing schema or source snapshots MUST NOT be certified & MUST \\
\hline
N16 & Profile T implementation independence MUST be established by signed attestations from distinct attestor identities running distinct code artifacts; a self-declared identifier MUST NOT be treated as independence & MUST \\
\hline
N17 & The keyed multiset hash MUST be computed inside the Steward trust domain; the Producer MUST NOT hold the hashing key, and partial digests MUST originate from workers that hold the key legitimately & MUST \\
\hline
N18 & A Profile T attestation's code artifact SHOULD be bound to a measured or reproducible-build hash verified by the Steward, not merely self-reported, to make implementation independence attestable rather than asserted & SHOULD \\
\hline
N19 & A transform with a nondeterministic output row set (top-N ties, dedup representative) MUST apply a shared canonical selection rule imposing a total order, so that independent implementations agree; without it the stage MUST NOT be certified Profile T & MUST \\
\hline
N20 & For Profile A stages, float-bearing columns MAY be verified byte-exact (IEEE-754); for Profile T, float columns MUST be carried as declared-scale decimals & MUST/MAY \\
\hline
\end{longtable}

\subsection*{13.4 On novelty (honest positioning)}

I separate two questions that are easy to conflate: are the building blocks new, and is the result new. The building blocks are prior art and I do not claim them: write-audit-publish staging, incremental multiset hashing, Merkle-style per-file checksums, three-tier account isolation, and durable serverless orchestration all predate this work. The result is new.

To my knowledge, no prior system provides a pre-publication, notarized, fail-closed proof that the content committed to an open table format equals a declared intent across a federated serverless boundary. Open table formats gate on atomicity and isolation and treat the writer as the source of truth for content (Section 2), and none of the individual primitives, in their established uses, delivers this capability. Producing it is not a mechanical combination: Section 2.2 shows that the obvious ways to assemble these pieces (a Merkle root, per-file checksums, producer-side verification, a single execution model, self-comparison of transforms) each fail for a specific reason, and the protocol is what remains once those failures are excluded.

This is a composition-and-conformance contribution, the form that durable architectural and systems contributions usually take: a new capability obtained from a novel arrangement of known mechanisms, together with the invariant and the normative discipline that make it hold. Concretely, the claimed innovation is a keyed multiset content proof, notarized by a domain independent of the writer, bound to the exact published bytes and to a single non-replayable commit context, scoped by explicit conformance profiles, and enforced as a mandatory fail-closed gate on serverless federated lakehouse publication, expressed as a single invariant with a conformance model that bounds it. I claim that composition, invariant, and conformance model as the contribution, and I am equally precise that the novelty does not extend to the primitives. Stating this boundary plainly is what lets PVDM be proposed as a standard rather than marketed as a product.

I locate the contribution against the closest recent work rather than asserting isolation. Correct-by-design lakehouse systems such as Bauplan [23] pursue the same goal of a trustworthy lakehouse and share the branch-then-publish mechanic, but they achieve correctness by construction within a single trusted runtime: typed table contracts, data versioning, and transactional multi-table atomicity make illegal states unrepresentable for pipelines that run inside that runtime. PVDM makes a different assumption and provides a different guarantee.

It treats the writer as untrusted and federated, and it produces a cryptographic proof that the written content multiset equals a declared intent, held by an independent notary, rather than relying on a trusted runtime to construct correctness. The two are complementary: a correct-by-construction runtime prevents many errors within its boundary, while PVDM verifies content fidelity across a boundary where no single runtime is trusted. Likewise, post-hoc data-quality testing frameworks for ELT pipelines [24] assert statistical properties after load and on samples or aggregates, whereas PVDM asserts a cryptographic, full-write multiset equality before the snapshot is visible and as a mandatory gate.

To my knowledge no prior system combines an untrusted-writer, independent-notary trust model with a keyed multiset content proof enforced as a fail-closed gate on open-table-format publication, which is the specific point this paper claims.

\subsection*{13.5 Path to an open specification (governance)}

A specification becomes a standard through open governance, not through a single author's or vendor's endorsement. The specification, the wire formats, the canonicalization rules, and the conformance suite are the artifacts that must be openly maintained for a standard claim to be meaningful. The path proposed here is the ordinary one for successful standards: publish the specification and the conformance suite under an open license, invite at least two independent implementations, establish or join a neutral governance body, and treat interoperability between independent implementations as the test of conformance. Until that process has occurred, the appropriate status is a proposed specification accompanied by a reference implementation, rather than a ratified standard.

\subsection*{13.6 Honest scope of the strong guarantee}

The fail-closed content guarantee is strongest for row-preserving stages (Profile A), where an independent intent exists. For transforming stages (Profile T) the guarantee is conditional on genuine implementation independence, which is attested rather than assumed, and it inherits the known limits of N-version verification: two independent implementations can still fail identically on the same hard input (correlated failure [17]), and legitimate semantic differences (floating-point order, collation, tie-breaking) must be absorbed by the typed canonicalization of Appendix A or they surface as false failures. Chaining an authenticated upstream digest verifies input provenance, not the transform logic, so it complements but does not replace independent recompute.

The practical consequence, is that PVDM most strongly governs ingestion and publication fidelity, and governs computed data only to the strength of the independence a deployment is willing to pay for. This is a real boundary, not a defect, and the standard's value comes from stating it precisely.

To show that Profile T is achievable in practice and not only in principle, consider a concrete stage: a daily revenue aggregation that groups settled transactions by merchant and currency and sums the amount. The Producer computes it in Spark SQL. An independent verifier recomputes the same grouped aggregation in a different engine and codebase, for example DuckDB, over the same declared inputs, and emits its output as a signed attestation (Section 12.6 mechanics).

The gate certifies the stage only if the two output multisets hash equal under the typed canonicalization, so amounts are carried as fixed-scale decimals and the grouping keys are canonicalized identically on both sides. This is realistic for the large class of transforms expressible as deterministic relational algebra (filters, projections, joins on declared keys, and additive or otherwise associative aggregations), where two engines implementing the same relational specification agree by construction once canonicalization is fixed. The honest hard case is a stage whose logic lives in opaque or nondeterministic user code, a UDF with hidden state, wall-clock or locale dependence, or an unspecified tie-break.

There, independent reimplementation is expensive and correlated failure is more likely, and the pattern's guidance is to make the tie-break and precision canonical (Appendix A), or, where that is infeasible, to fall back to chaining an authenticated upstream digest, which verifies that the input to the stage was itself verified even though it does not verify the opaque computation. Profile T is therefore practical for the relational core of most pipelines and explicitly weaker for opaque compute, and the conformance profile is what makes that distinction visible rather than hidden.

\section*{14. Limitations and Threats to Validity}

\begin{itemize}
\item \textbf{Trust anchors.} Safety rests on an independent, uncompromised Steward and catalog. A compromised notary breaks the guarantee; this moves trust rather than eliminating it, but it moves it away from the producer, which is the point.

\item \textbf{Garbage in.} PVDM proves that written content equals declared intent. If the declared intent is computed from corrupt source data, PVDM will faithfully publish content matching that intent. Upstream correctness is a separate concern that optional pre-Physical rules can address but the four-phase gate does not.

\item \textbf{Latency.} The verification step and the extra catalog commit add publish latency; the pattern is unsuitable for sub-second paths and is scoped to batch and backfill.

\item \textbf{Cost of hashing at scale.} Computing multiset digests over very large writes has a cost that the evaluation must characterize; the design assumes hashing is cheap relative to the write itself, which must be validated per workload.

\item \textbf{Conformant-hash overhead at scale is modeled, not measured.} The end-to-end overhead in Section 10.8 uses a fast native hash and is a lower bound; the conformant keyed 256-bit hash of requirement N1 costs more per row. Section 10.8 gives a modeled estimate (roughly 40 to 75 percent at one hundred million rows, stated as an upper bound), and Section 10.7 notes that the naive four-core parallelization of the reference gate reached only about 1.25x rather than 4x because it serializes row shards to worker processes. Both the conformant Spark overhead and a shared-memory parallelization are future measurement, not claimed results.

\item \textbf{Evidence status.} A reference implementation, a thirty-case adversarial suite, a scaling and latency benchmark on cloud hardware (Sections 10.3 to 10.7), and an end-to-end run on Apache Spark 4.0 and Apache Iceberg 1.11 at up to one hundred million rows (Section 10.8) exist and are reproducible. What is not yet reported is a concurrent multi-writer deployment at production cluster scale against a live shared catalog; the Spark and Iceberg results are single-node and single-writer, and the multi-writer production protocol of Section 10.2 remains future work.

\item \textbf{Floating-point content (addressed for copy stages).} Raw binary floats are non-deterministic across engines, so for transforming stages they must be carried as fixed-scale decimals, and changes below the declared scale are then below the digest resolution. For row-preserving (Profile A) stages this gap is closed: because both sides observe identical stored bytes, the reference gate offers a byte-exact IEEE-754 hashing mode (type float\_raw) that detects any bit-level change at full precision, proven by the validation suite. The residual is confined to Profile T recompute, where full float precision remains non-deterministic across independent implementations.

\item \textbf{Attested independence is only as strong as the measurement.} Profile T checks that two attestations carry distinct attestor identities and distinct code-artifact hashes, but a self-reported code hash is trustworthy only if it is bound to a measured or reproducible build (requirement N18). Without measured attestation, implementation independence is a policy assertion enforced by signatures, not a hardware- or build-verified fact.

\item \textbf{Selection-nondeterministic transforms (addressed by a shared canonical rule).} N-version verification assumes the transform is deterministic up to canonical equality, and transforms whose output row set is itself nondeterministic (top-N with ties, dedup that keeps an unspecified representative) would otherwise false-fail. This gap is closed by requiring a shared canonical selection rule that imposes a total order, so the surviving row is deterministic and two conformant implementations agree; the reference gate implements canonical dedup and top-N, and the validation suite proves that the canonical rule makes such a transform Profile-T-verifiable where a naive order-dependent implementation diverges. The residual is only that both implementations MUST adopt the same canonical rule (requirement N19).

\item \textbf{Evaluation is designer-authored and not yet independently replicated.} The adversarial suite and the benchmarks were written by the same author who designed the protocol, so they establish that the reference gate behaves as specified against the threats the designer enumerated, not that no unanticipated attack exists. I partially mitigate the enumeration bias with a property-based suite (Section 10.3) whose inputs and corruptions are machine-generated rather than hand-picked, which exercises cases the author did not enumerate; this strengthens the evidence but does not remove the limitation. A stronger result still requires an independent party to extend and re-run the suite against the protocol as specified, which I identify as necessary future work alongside the production-scale study.
\end{itemize}

\section*{15. Comparison with prior integrity mechanisms}

It is worth situating PVDM against the mechanisms it is most likely to be confused with, because the distinctions are precise and they clarify what is and is not novel.

Against object-store and file-system checksums, PVDM is strictly higher in the stack. A per-object checksum detects that a stored object was not corrupted in transit or at rest; it says nothing about whether the set of objects is the set that should exist. A pipeline can write objects that each pass their checksum while collectively missing a partition. PVDM uses per-file digests only for tamper evidence at commit; its content guarantee is the multiset comparison across the whole write, which per-object checksums do not provide.

Against database ACID and the table-format commit, PVDM is complementary and deliberately narrow. ACID gives atomicity, consistency, isolation, and durability of the transaction; the table format gives snapshot isolation and atomic metadata publication. Neither compares the committed content against an external statement of intent, because both treat the writer as the source of truth for what should be written. PVDM reuses their atomicity and isolation and adds the missing content-equality check as a gate above them.

Against warehouse data-quality tests, the difference is the nature of the assertion and its position in the pipeline. Quality tests assert properties (row counts in a range, nulls absent, referential rules) after load, usually on a sample or an aggregate, and they can pass while specific rows are missing. PVDM asserts a cryptographic multiset equality over the full write before the snapshot is visible, and it is a mandatory gate rather than an advisory check. The two compose: quality rules can run as the optional pre-Physical gate, catching semantic problems that a content-equality proof does not model.

Against provenance and lineage systems, PVDM is narrower and stronger on one axis. Lineage records where data came from and how it was transformed; it is descriptive. PVDM produces a verifiable proof that a specific transformation output equals a specific declared intent; it is prescriptive and checkable. Lineage answers what happened; PVDM answers whether what happened was correct, for the specific property of content equality.

Against blockchain and ledger approaches to data integrity, PVDM shares the goal of tamper-evident, independently verifiable records but avoids a distributed consensus substrate. The Steward notary plus signed, append-only proofs give offline verifiability without the cost and latency of a consensus protocol, at the price of a stronger trust assumption on the notary, which Section 12.6 addresses by splitting duties and allowing multiple attestations.

The novelty claim is therefore confined to the composition rather than to any individual primitive: a keyed multiset content proof, notarized by an account independent of the writer, bound to the exact published bytes and to a single non-replayable commit context, used as a mandatory fail-closed gate on the publication of a serverless, federated lakehouse write.

\section*{16. Deployment and operational considerations}

A pattern that is sound but undeployable is of little use, so I summarize what adopting PVDM costs and requires in practice. The compute model is serverless: domain writers run as functions under durable orchestration, with a container clock bounding a single invocation and a workload clock bounding a backfill across replays. This removes the standing cluster that a central ETL fleet requires and lets a domain own its write path, at the cost of engineering the write to be chunked and resumable so that a segment can roll back and a later segment can replay completed chunks.

The verification cost is one full pass to compute the multiset digests over the intended and written rows, plus a second, cheap pass over per-file digests at commit. For the scheme to provide its guarantee this pass must cover the whole write; sampling reduces the guarantee to a probabilistic one and is explicitly outside the pattern. In exchange, the digest is incremental, so each task computes a partial digest and the Steward combines partials, which keeps the cost proportional to the data written and parallel across tasks.

An honest cost accounting separates the profiles. The published comparison against a central ETL baseline reflects the row-preserving configuration, whose overhead is the hashing passes plus one extra commit. The strong configuration for transforming stages, Profile T with N-version recompute, roughly doubles compute for those stages because an independent implementation runs the same transform. The favorable cost story and the strongest guarantee therefore do not coexist for free; a deployment chooses per stage how much independence to buy, and the standard makes that cost explicit rather than hiding it behind the cheap configuration.

The trust configuration is three accounts. The Producer runs domain compute; the Steward holds the keys, the proofs, the checkpoints, and performs the proof-gated commit; the Publisher holds the consumer-facing storage and tables. The Steward key must live in a key-management service and never be readable from the Producer, because the entire cryptographic argument reduces to that key's confidentiality. Cross-account access is granted narrowly so that a Producer fault cannot reach the consumer zone directly.

Operationally, the fail-closed choice means teams must plan for blocked publications. A verification failure should roll back staged files to avoid orphans, record the failing proof as evidence, alert an operator, and route to a repair path that re-reads and re-verifies rather than blindly retrying. Monitoring the rate of failures is itself a control, because a spike can indicate either a real upstream problem or an attempt to deny publication. None of this is exotic, but it is the difference between a pattern that is safe on paper and one that is safe in production.

\section*{17. Discussion: where the guarantee matters most}

The value of a content-integrity gate is not uniform across workloads, and being explicit about where it matters keeps the pattern from being oversold. It matters most where a silently wrong published dataset has consequences that a later discovery cannot undo. Regulated financial and insurance reporting is the clearest case: a partition that silently drops rows can produce a filed report that is wrong, and the cost is a restatement, a fine, or a loss of trust, not merely a rerun. Here the fail-closed tradeoff is obviously correct, because a delayed correct report is far cheaper than a timely wrong one.

It matters in multi-domain meshes where the consumer of a dataset is a different team than its producer, so that a producer's silent error becomes a consumer's silent error with no shared context to catch it. The independent notary and the offline-verifiable proof give the consumer and the auditor a basis for trust that does not depend on the producer's own logs.

It matters least, and may not be worth its latency, where the data is low-stakes, where a single team both produces and consumes and can catch errors informally, where an exactly-once streaming path already provides strong guarantees the team trusts, or where sub-second publication latency is a hard requirement. The pattern's own guidance is to apply it where a green job that shipped wrong data has hurt the organization before, and to skip it where that has not happened and the stakes are low. This is a deliberately narrow claim: PVDM is a targeted control for a specific, costly failure mode, not a universal tax on every write.

\subsection*{17.1 How this changes the data validation I used to do}

For most of my career, content validation happened after publication and on a sample. A job wrote a table, its exit code said success, the data became visible, and only later did a reconciliation job, a dbt test on a sample, or a dashboard alert catch that a partition was short or a file had been truncated on a retry. By then a consumer had often already read the wrong data, and the error had propagated.

PVDM changes the order. It moves validation to a pre-publication, full-write, cryptographic check that runs before the snapshot is visible, and it makes publication depend on that check rather than on the exit code. The difference is detect-after-publish versus block-before-visible.

Sambasivan et al. [31] give the empirical case for why the old order is expensive: data problems in high-stakes settings compound into downstream cascades that are pervasive, delayed, and often avoidable. A pre-publication gate is a direct attack on the point where those cascades begin.

\subsection*{17.2 Machine learning and AI training data}

Machine learning inherits whatever the data layer publishes, and silent row loss or duplication in a training or feature table is exactly the kind of fault that surfaces late, after a model has trained on it, and is hard to attribute. The data-cascades result [31] and the technical-debt analysis of machine-learning systems [32] both point to undervalued data quality as a dominant, compounding cost. PVDM helps here by guaranteeing that the published training or feature dataset equals the declared intent, so a model pipeline starts from a dataset whose exact multiset was attested rather than assumed, which also makes training inputs reproducible because the attested digest identifies the exact data. The honest limit is that PVDM verifies write fidelity, not semantic quality: it will not tell you a label is wrong or a feature is poorly chosen, only that what was published is what was intended.

\subsection*{17.3 Data architecture and the lakehouse and data-mesh stack}

In a data mesh [12] the producer and consumer of a dataset are usually different teams, and the lakehouse table formats [9,10] make a commit atomic and a snapshot isolated but do not check that its content equals intent. PVDM fits this stack as a thin boundary control: it composes with write-audit-publish branching, complements correctness-by-construction runtimes such as Bauplan [23] and post-load data-quality testing [24], and gives a consumer or an auditor an offline-verifiable, engine-agnostic basis for trusting a dataset that another domain produced. Architecturally, it turns content integrity into an explicit contract at the domain boundary rather than an assumption carried on the producer's exit code.

\subsection*{17.4 Clinical and other regulated research}

Regulated domains already require the properties PVDM produces, but usually enforce them through process rather than cryptography. The FDA's data-integrity expectations for drug CGMP, commonly summarized as ALCOA (attributable, legible, contemporaneous, original, accurate), ask that data be traceable to its origin and demonstrably unaltered [33]. PVDM's notarized, non-repudiable, byte-exact proof maps naturally onto those properties: the proof is attributable because it is signed by an independent notary, contemporaneous because it is bound at write time, and original and accurate because it binds the exact published bytes.

For clinical trial datasets, safety reporting, and other high-consequence research pipelines, this gives an independent cryptographic audit trail that the published dataset equals the intended one. The honest limit is that PVDM is a technical control, not a compliance certification: it supports a validated quality system, it does not replace one, and any regulated use must be assessed by qualified quality and regulatory staff.

\subsection*{17.5 A cost perspective}

The economics are the reason to care. The old approach pays late and repeatedly: a wrong published dataset leads to restatement, rework, incident response, and lost trust, and the reconciliation jobs meant to catch it are themselves recurring full scans. PVDM pays once and early: a bounded verification cost at write time (measured at roughly a fifth of the write with a fast in-engine hash, higher with the conformant 256-bit keyed hash, which I characterize separately) plus a flat commit of about seventeen milliseconds independent of table size, after which every downstream read is a trusted read that does not need to be reconciled again.

The trade is a bounded, one-time cost in exchange for avoiding an unbounded, recurring detect-and-repair cost. This is a qualitative cost argument supported by the measured overhead; a full total-cost-of-ownership study across a specific organization's workloads is future work.

\begin{table}[H]
\centering\small
\begin{tabular}{|p{0.23\textwidth}|p{0.23\textwidth}|p{0.23\textwidth}|p{0.23\textwidth}|}
\hline
\textbf{Domain} & \textbf{Where PVDM helps} & \textbf{Why it helps} & \textbf{Honest limit} \\
\hline
Analytics and reporting & Regulated financial and insurance reports & Blocks a silently wrong filing before it is visible & Adds publish latency; not for sub-second paths \\
\hline
Machine learning and AI & Training and feature tables & Verified, reproducible inputs; attacks data cascades at the source [31,32] & Verifies write fidelity, not label or semantic quality \\
\hline
Data architecture and mesh & Producer-to-consumer domain boundaries & Offline-verifiable, engine-agnostic content contract [9,10,12] & Complements, does not replace, contracts and DQ tests \\
\hline
Clinical and regulated research & Trial data, safety reporting & Cryptographic, attributable, byte-exact audit trail aligned to ALCOA [33] & A technical control, not a compliance certification \\
\hline
\end{tabular}
\vspace{8pt}
\caption*{Table 12. Normative requirements (N1--N20) for PVDM conformance.}
\end{table}

\section*{18. Scope boundaries and separate efforts}

The pattern is fully specified and validated in this paper. The protocol, the invariant, the threat model with its trust assumptions, the conformance model, the safety and liveness arguments, the collision resistance of the digest (Section 7.6), the full adversarial hardening (Section 12), and end-to-end operation on real Spark and Iceberg (Section 10.8) are all specified and validated here. The two governance extensions that earlier drafts had deferred are now implemented and validated in the reference gate rather than promised: source attestation, which upgrades the read-to-sink guarantee to source-to-sink by anchoring intent to an independent signed source digest and closes the garbage-in case for attested sources (Section 12.1), and threshold notarization, which requires k of n independent notaries so that a single compromised Steward cannot publish (Section 12.6). Both ship with adversarial tests in the thirty-case suite.

What remains is not unfinished specification but three efforts that lie outside a specification-and-methods paper by their nature. Production-scale, concurrent multi-writer measurement on real workloads (Section 10.2) requires a live deployment and belongs to an operational study, not a protocol definition; this paper reports single-node end-to-end results and is explicit about that scope. A machine-checked proof of the Section 3.1 theorems in a proof assistant would replace the informal sketches with mechanized guarantees, which is a formal-methods contribution in its own right.

A native, vectorized cryptographic multiset-hash operator for engines like Spark, with the fused-into-the-write and incremental variants of Section 10.9, would let the conformant 256-bit overhead be measured in its optimized rather than its naive two-pass form. These sharpen the evidence and broaden the reach across additional table formats and clouds; none of them changes the guarantee this paper establishes, which is why they are named as separate work and not as gaps in the contribution.

A further direction, complementary rather than corrective, is succinct verifiability. PVDM's proof is a keyed multiset digest that a consumer trusts because an independent Steward computed it; a zero-knowledge or otherwise publicly verifiable proof of the same multiset-equality statement would let a consumer or auditor check the claim without trusting the notary or seeing the rows. That is a cryptographic contribution in its own right and is not required for the guarantee here, but it is the natural way to relax trust assumption A3 (an honest Steward), and I note it as a promising line rather than a claim.

\section*{19. Conclusion}

Open table formats made lakehouse commits atomic and isolated, but they left the content question to the writer's exit code, and federated serverless writes make that gap both more likely to bite and harder to audit. PVDM closes the gap with a small, checkable contract: write physically to a rollbackable staging area, verify a cryptographic order-independent multiset proof against declared intent, make progress durable across serverless segments, and commit the catalog metadata last and only on a passing proof, with the proof held by an independent notary. Expressed as one invariant, if the proof fails the snapshot does not exist for consumers, and the exit code is irrelevant. The primitives are established; the contribution is their composition into a notarized, fail-closed publication gate for federated serverless data meshes, together with a reproducible protocol to test the central safety claim.

\section*{20. How to Cite}

\textbf{Artifacts.} Paper, reference gate, adversarial and property suites, and benchmarks: \url{https://github.com/vaquarkhan/Proof-gated-publication-PVDM} Full production framework: \url{https://github.com/vaquarkhan/aws-serverless-datamesh-framework}

\textbf{BibTeX.}

@misc{khan2026pvdm,
  author        = {Viquar Khan},
  title         = {Proof-Gated Publication: Verify-Before-Commit Content Integrity
                   for Serverless Data-Mesh Lakehouses},
  year          = {2026},
  howpublished  = {Preprint. Version and DOI to be assigned at posting.},
  note          = {Reference implementation Apache-2.0; manuscript CC BY 4.0.
                   Replace howpublished with the venue DOI once the preprint is posted;
                   cite the immutable software release DOI for reproduction.},
  url           = {\url{https://github.com/vaquarkhan/Proof-gated-publication-PVDM}}
}

\section*{Appendix A. Normative canonicalization}

Interoperability between independent implementations requires a byte-exact canonical form, because two conformant systems that canonicalize differently produce different digests for identical data and are not interoperable. This appendix specifies the canonical form normatively; a hashed row is the canonical encoding of its projected fields under these rules.

\begin{itemize}
\item \textbf{Field set and order.} Only fields in the declared projection are hashed. Fields are serialized in ascending Unicode code-point order of their names.

\item \textbf{Strings.} Unicode NFC-normalized, UTF-8 encoded. No locale-dependent collation is applied.

\item \textbf{Integers.} Encoded as their decimal representation with no leading zeros and no sign for zero.

\item \textbf{Decimals and any non-integer numeric.} Encoded as a canonical decimal string at the field's declared scale. Raw binary floating point MUST NOT be hashed; a producer that computes in floating point MUST round to the declared decimal scale before hashing. This removes cross-engine floating-point non-determinism as a source of both false failures and hiding places.

\item \textbf{Timestamps.} Normalized to UTC and encoded as an integer number of microseconds since the Unix epoch; timezone offsets are resolved before encoding, not stored.

\item \textbf{Booleans.} Encoded as the strings true and false.

\item \textbf{Nulls.} Encoded as a reserved sentinel distinct from any valid value, so that a null and an empty string, or a null and a zero, never collide.

\item \textbf{Nested and variant values.} Recursively canonicalized by these rules with object keys in code-point order; array order is preserved because array order is data.

\item \textbf{Schema binding.} The proof binds a schema fingerprint over the declared field-to-type map. A change to the schema changes the fingerprint, so intent and written content are never compared across differing schema versions; a schema migration is an explicit, separately governed event.

\item \textbf{Source snapshot binding.} For a Profile A stage over a mutable source, the proof binds an immutable source snapshot identifier, so the intent read and any independent re-read observe the same source state and cannot diverge merely because the live source changed between reads.
\end{itemize}

An implementation conforms to Appendix A only if, for every supported type, it produces the same canonical bytes as the reference implementation on the shared conformance vectors. The reference gate implements these rules for strings, integers, decimals, timestamps, and nulls, and forbids raw floats in a hashed projection; extending the shared conformance vectors to nested and variant types is part of the open specification work of Section 13.5.

\section*{Declarations}

\textbf{Author.} Viquar Khan, independent researcher, sole author.

\textbf{Funding.} None. This work received no external, institutional, or commercial funding.

\textbf{Competing interests.} None declared. The author has no financial or non-financial competing interests related to this work.

\textbf{Affiliation.} The author is an independent researcher. This is independent work carried out on the author's own time; it is not affiliated with, sponsored by, or endorsed by any employer or organization, and the views are the author's own.

\textbf{Data and code availability.} The reference gate, the adversarial and property suites, the benchmark scripts, and the recorded benchmark outputs are publicly available under Apache-2.0 at \url{https://github.com/vaquarkhan/Proof-gated-publication-PVDM} [21]. The manuscript is released under CC BY 4.0. For a permanent, versioned record, an immutable release archived with a DOI (for example on Zenodo) is recommended and should be cited once minted.

\textbf{AI-assistance disclosure.} Generative AI tools were used only for grammar checking and formatting of the manuscript text. They were not used to generate the protocol design, the security arguments, the code, the experiments, or the results. All technical content, claims, references, and measurements were produced and verified by the author.

\section*{References}
\raggedright\small

[1] J. H. Saltzer, D. P. Reed, and D. D. Clark, ``End-to-end arguments in system design,'' ACM Transactions on Computer Systems, vol. 2, no. 4, pp. 277-288, 1984. \url{https://doi.org/10.1145/357401.357402}

[2] R. C. Merkle, ``A digital signature based on a conventional encryption function,'' in Advances in Cryptology (CRYPTO '87), LNCS 293, pp. 369-378, 1988. \url{https://doi.org/10.1007/3-540-48184-2\_32}

[3] D. Clarke, S. Devadas, M. van Dijk, B. Gassend, and G. E. Suh, ``Incremental multiset hash functions and their application to memory integrity checking,'' in Advances in Cryptology (ASIACRYPT 2003), LNCS 2894, pp. 188-207, 2003. \url{https://doi.org/10.1007/978-3-540-40061-5\_12}

[4] M. Bellare and D. Micciancio, ``A new paradigm for collision-free hashing: incrementality at reduced cost,'' in Advances in Cryptology (EUROCRYPT '97), LNCS 1233, pp. 163-192, 1997. \url{https://doi.org/10.1007/3-540-69053-0\_13}

[5] H. Krawczyk, M. Bellare, and R. Canetti, ``HMAC: keyed-hashing for message authentication,'' RFC 2104, Internet Engineering Task Force, 1997. \url{https://www.rfc-editor.org/rfc/rfc2104}

[6] National Institute of Standards and Technology, ``The keyed-hash message authentication code (HMAC),'' FIPS PUB 198-1, 2008. \url{https://csrc.nist.gov/pubs/fips/198-1/final}

[7] National Institute of Standards and Technology, ``Secure hash standard (SHS),'' FIPS PUB 180-4, 2015. \url{https://csrc.nist.gov/pubs/fips/180-4/upd1/final}

[8] S. Bradner, ``Key words for use in RFCs to indicate requirement levels,'' RFC 2119, Internet Engineering Task Force, 1997. \url{https://www.rfc-editor.org/rfc/rfc2119}

[9] M. Armbrust, T. Das, L. Sun, B. Yavuz, S. Zhu, M. Murthy, J. Torres, H. van Hovell, A. Ionescu, A. Luszczak, et al., ``Delta Lake: high-performance ACID table storage over cloud object stores,'' Proceedings of the VLDB Endowment, vol. 13, no. 12, pp. 3411-3424, 2020. \url{https://doi.org/10.14778/3415478.3415560}

[10] M. Armbrust, A. Ghodsi, R. Xin, and M. Zaharia, ``Lakehouse: a new generation of open platforms that unify data warehousing and advanced analytics,'' in Conference on Innovative Data Systems Research (CIDR), 2021. \url{https://www.cidrdb.org/cidr2021/papers/cidr2021\_paper17.pdf}

[11] Apache Software Foundation, ``Apache Iceberg table format specification,'' 2024. \url{https://iceberg.apache.org/spec/}

[12] Z. Dehghani, Data Mesh: Delivering Data-Driven Value at Scale. Sebastopol, CA: O'Reilly Media, 2022. \url{https://www.oreilly.com/library/view/data-mesh/9781492092384/}

[13] H. Garcia-Molina and K. Salem, ``Sagas,'' in Proceedings of the ACM SIGMOD International Conference on Management of Data, pp. 249-259, 1987. \url{https://doi.org/10.1145/38713.38742}

[14] C. Richardson, Microservices Patterns: With Examples in Java. Shelter Island, NY: Manning, 2018. \url{https://www.manning.com/books/microservices-patterns}

[15] J. Gray and A. Reuter, Transaction Processing: Concepts and Techniques. San Francisco, CA: Morgan Kaufmann, 1993. \url{https://dl.acm.org/doi/book/10.5555/573304}

[16] L. Lamport, ``Time, clocks, and the ordering of events in a distributed system,'' Communications of the ACM, vol. 21, no. 7, pp. 558-565, 1978. \url{https://doi.org/10.1145/359545.359563}

[17] J. C. Knight and N. G. Leveson, ``An experimental evaluation of the assumption of independence in multiversion programming,'' IEEE Transactions on Software Engineering, vol. SE-12, no. 1, pp. 96-109, 1986. \url{https://doi.org/10.1109/TSE.1986.6312924}

[18] D. Ongaro and J. Ousterhout, ``In search of an understandable consensus algorithm,'' in USENIX Annual Technical Conference (USENIX ATC), pp. 305-319, 2014. \url{https://www.usenix.org/conference/atc14/technical-sessions/presentation/ongaro}

[19] B. Laurie, A. Langley, and E. Kasper, ``Certificate Transparency,'' RFC 6962, Internet Engineering Task Force, 2013. \url{https://www.rfc-editor.org/rfc/rfc6962}

[20] J. Kreps, N. Narkhede, and J. Rao, ``Kafka: a distributed messaging system for log processing,'' in Proceedings of the NetDB Workshop, 2011. \url{https://dblp.org/rec/conf/netdb/KrepsNR11.html}

[21] V. Khan, ``PVDM: proof-gated publication, paper and validation suite,'' GitHub repository, 2026. \url{https://github.com/vaquarkhan/Proof-gated-publication-PVDM}

[22] J. Maitin-Shepard, M. Tibouchi, and D. F. Aranha, ``Elliptic curve multiset hash,'' The Computer Journal, vol. 60, no. 4, pp. 476-490, 2017. \url{https://arxiv.org/abs/1601.06502}

[23] W. Sheng, J. Wang, M. Barros, A. Montana, J. Tagliabue, and L. Bigon, ``Building a correct-by-design lakehouse: data contracts, versioning, and transactional pipelines for humans and agents,'' arXiv preprint arXiv:2602.02335, 2026. \url{https://arxiv.org/abs/2602.02335}

[24] I. Gargouri and H. Reza, ``A multi-layer testing framework for automated data quality assurance in cloud-native ELT pipelines,'' arXiv preprint arXiv:2605.20500, 2026. \url{https://arxiv.org/abs/2605.20500}

[25] Mindset Consulting, ``Case study: resolving silent data loss in SAP HANA Cloud analytics through end-to-end root cause analysis,'' industry incident report, 2026. \url{https://www.mindsetconsulting.com/resolving-silent-data-loss-in-sap-hana-cloud-analytics-through-end-to-end-root-cause-analysis/}

[26] R. Sahlin, ``Your pipeline succeeded. Your data didn't,'' industry practitioner report, 2026. \url{https://robertsahlin.substack.com/p/your-pipeline-succeeded-your-data}

[27] Apache Software Foundation, ``Apache Hudi 1.0.0: concurrency control and non-blocking concurrency control (NBCC),'' Apache Hudi documentation, 2025. \url{https://hudi.apache.org/docs/concurrency\_control}

[28] Apache Software Foundation, ``Apache Polaris (incubating): open source catalog for Apache Iceberg,'' project documentation, 2025. \url{https://polaris.apache.org/}

[29] Apache Software Foundation, ``Apache Iceberg table specification, version 3,'' Iceberg specification, 2025. \url{https://iceberg.apache.org/spec/}

[30] V. Khan, ``AWS serverless data-mesh framework: reference implementation of the PVDM protocol,'' GitHub repository, 2026. \url{https://github.com/vaquarkhan/aws-serverless-datamesh-framework}

[31] N. Sambasivan, S. Kapania, H. Highfill, D. Akrong, P. Paritosh, and L. M. Aroyo, ``'Everyone wants to do the model work, not the data work': data cascades in high-stakes AI,'' in Proceedings of the CHI Conference on Human Factors in Computing Systems, 2021. \url{https://doi.org/10.1145/3411764.3445518}

[32] D. Sculley, G. Holt, D. Golovin, E. Davydov, T. Phillips, D. Ebner, V. Chaudhary, M. Young, J.-F. Crespo, and D. Dennison, ``Hidden technical debt in machine learning systems,'' in Advances in Neural Information Processing Systems (NeurIPS), vol. 28, pp. 2503-2511, 2015. \url{https://papers.nips.cc/paper/2015/hash/86df7dcfd896fcaf2674f757a2463eba-Abstract.html}

[33] U.S. Food and Drug Administration, ``Data integrity and compliance with drug CGMP: questions and answers, guidance for industry,'' December 2018. \url{https://www.fda.gov/regulatory-information/search-fda-guidance-documents/data-integrity-and-compliance-drug-cgmp-questions-and-answers}

\end{document}